\documentclass[epj,nopacs]{svjour}
\usepackage{lineno}
\usepackage{amsmath}
\usepackage{amssymb}
\usepackage{epsfig}
\usepackage{color}
\usepackage{vruler}
\usepackage{rotating}
\usepackage{cite}

\newcommand{\beq}  {\begin{equation}}
\newcommand{\eeq}  {\end{equation}}

\usepackage{ifpdf}
\ifpdf
\DeclareGraphicsExtensions{.pdf, .jpg, .tif}
\usepackage[%
  pdftitle={Ratios of Helicity Amplitudes for Exclusive $\rho^0$ Electroproduction 
on Transversely Polarized Proton},%
 pdfauthor={The HERMES Collaboration},%
  pdfsubject={HERMES amplitude ratios},%
  pdfstartview=FitH,%
  bookmarks=true,%
  bookmarksopen=true,%
  breaklinks=true,%
  colorlinks=true,%
  linkcolor=blue,anchorcolor=blue,%
  citecolor=blue,filecolor=blue,%
  menucolor=blue,pagecolor=blue,%
  urlcolor=blue]{hyperref}
\else
\DeclareGraphicsExtensions{.eps, .jpg}
\usepackage[%
  breaklinks=true,%
  colorlinks=true,%
  linkcolor=blue,anchorcolor=blue,%
  citecolor=blue,filecolor=blue,%
  menucolor=blue,pagecolor=blue,%
  urlcolor=blue]{hyperref}
 \fi
\begin{document}
\hugehead

\title{
  Ratios of helicity amplitudes for exclusive $\rho^0$  electroproduction on transversely polarized protons
}

\author{ 
The HERMES Collaboration \medskip \\
A.~Airapetian$^{14,17}$,
N.~Akopov$^{29}$,
Z.~Akopov$^{6}$,
E.C.~Aschenauer$^{7}$,
W.~Augustyniak$^{28}$,
S.~Belostotski$^{20}$,
H.P.~Blok$^{19,27}$,
A.~Borissov$^{6}$,
V.~Bryzgalov$^{21}$,
G.P.~Capitani$^{12}$,
G.~Ciullo$^{10,11}$,
M.~Contalbrigo$^{10}$,
W.~Deconinck$^{6}$,
R.~De~Leo$^{2}$,
E.~De~Sanctis$^{12}$,
M.~D\"uren$^{14}$,
G.~Elbakian$^{29}$,
F.~Ellinghaus$^{5}$,
L.~Felawka$^{25}$,
S.~Frullani$^{23,24}$,\footnote{deceased}
F.~Garibaldi$^{23,24}$,
G.~Gavrilov$^{6,20,25}$,
V.~Gharibyan$^{29}$,
S.V.~Goloskokov$^{8}$,
Y.~Holler$^{6}$,
A.~Ivanilov$^{21}$,
H.E.~Jackson$^{1}$,
S.~Joosten$^{13,16}$,
R.~Kaiser$^{15}$,
G.~Karyan$^{6,29}$,
A.~Kisselev$^{20}$,
V.~Korotkov$^{21}$,
V.~Kozlov$^{18}$,
P.~Kravchenko$^{9,20}$,
P.~Kroll$^{22}$,\footnote{Also at: Fachbereich Physik, Universit\"{a}t Wuppertal, 42096 Wuppertal, Germany}
L.~Lagamba$^{2}$,
L.~Lapik\'as$^{19}$,
I.~Lehmann$^{15}$,
P.~Lenisa$^{10,11}$,
W.~Lorenzon$^{17}$,
S.I.~Manaenkov$^{20}$,
B.~Marianski$^{28}$,
H.~Marukyan$^{29}$,
Y.~Miyachi$^{26}$,
A.~Movsisyan$^{10,29}$,
V.~Muccifora$^{12}$,
A.~Nass$^{9}$,
W.-D.~Nowak$^{7}$,
L.L.~Pappalardo$^{10,11}$,
A.~Petrosyan$^{29}$,
P.E.~Reimer$^{1}$,
A.R.~Reolon$^{12}$,
C.~Riedl$^{7,16}$,
K.~Rith$^{9}$,
G.~Rosner$^{15}$,
A.~Rostomyan$^{6}$,
D.~Ryckbosch$^{13}$,
Y.~Salomatin$^{21}$,\footnote{deceased}
A.~Sch\"afer$^{22}$,
G.~Schnell$^{3,4,13}$,
B.~Seitz$^{15}$,
T.-A.~Shibata$^{26}$,
V.~Shutov$^{8}$,
M.~Statera$^{10,11}$,
A.~Terkulov$^{18}$,
R.~Truty$^{16}$,
A.~Trzcinski$^{28}$,
M.~Tytgat$^{13}$,
Y.~Van~Haarlem$^{13}$,
C.~Van~Hulse$^{3,13}$,
D.~Veretennikov$^{20}$,
V.~Vikhrov$^{20}$,
I.~Vilardi$^{2}$,
S.~Yaschenko$^{6,9}$,
B.~Zihlmann$^{6}$,
P.~Zupranski$^{28}$
}

\institute{ 
$^1$Physics Division, Argonne National Laboratory, Argonne, Illinois 60439-4843, USA\\
$^2$Istituto Nazionale di Fisica Nucleare, Sezione di Bari, 70124 Bari, Italy\\
$^3$Department of Theoretical Physics, University of the Basque Country UPV/EHU, 48080 Bilbao, Spain\\
$^4$IKERBASQUE, Basque Foundation for Science, 48013 Bilbao, Spain\\
$^5$Nuclear Physics Laboratory, University of Colorado, Boulder, Colorado 80309-0390, USA\\
$^6$DESY, 22603 Hamburg, Germany\\
$^7$DESY, 15738 Zeuthen, Germany\\
$^8$Joint Institute for Nuclear Research, 141980 Dubna, Russia\\
$^9$Physikalisches Institut, Universit\"at Erlangen-N\"urnberg, 91058 Erlangen, Germany\\
$^{10}$Istituto Nazionale di Fisica Nucleare, Sezione di Ferrara, 44122 Ferrara, Italy\\
$^{11}$Dipartimento di Fisica e Scienze della Terra, Universit\`a di Ferrara, 44122 Ferrara, Italy\\
$^{12}$Istituto Nazionale di Fisica Nucleare, Laboratori Nazionali di Frascati, 00044 Frascati, Italy\\
$^{13}$Department of Physics and Astronomy, Ghent University, 9000 Gent, Belgium\\
$^{14}$II. Physikalisches Institut, Justus-Liebig Universit\"at Gie{\ss}en, 35392 Gie{\ss}en, Germany\\
$^{15}$SUPA, School of Physics and Astronomy, University of Glasgow, Glasgow G12 8QQ, United Kingdom\\
$^{16}$Department of Physics, University of Illinois, Urbana, Illinois 61801-3080, USA\\
$^{17}$Randall Laboratory of Physics, University of Michigan, Ann Arbor, Michigan 48109-1040, USA \\
$^{18}$Lebedev Physical Institute, 117924 Moscow, Russia\\
$^{19}$National Institute for Subatomic Physics (Nikhef), 1009 DB Amsterdam, The Netherlands\\
$^{20}$B.P. Konstantinov Petersburg Nuclear Physics Institute, Gatchina, 188300 Leningrad Region, Russia\\
$^{21}$Institute for High Energy Physics, Protvino, 142281 Moscow Region, Russia\\
$^{22}$Institut f\"ur Theoretische Physik, Universit\"at Regensburg, 93040 Regensburg, Germany\\
$^{23}$Istituto Nazionale di Fisica Nucleare, Sezione di Roma, Gruppo Collegato Sanit\`a, 00161 Roma, Italy\\
$^{24}$Istituto Superiore di Sanit\`a, 00161 Roma, Italy\\
$^{25}$TRIUMF, Vancouver, British Columbia V6T 2A3, Canada\\
$^{26}$Department of Physics, Tokyo Institute of Technology, Tokyo 152, Japan\\
$^{27}$Department of Physics and Astronomy, VU University, 1081 HV Amsterdam, The Netherlands\\
$^{28}$National Centre for Nuclear Research, 00-689 Warsaw, Poland\\
$^{29}$Yerevan Physics Institute, 375036 Yerevan, Armenia\\
}

\date{DESY Report 17-017  / Compiled: \today / Version: 5.0}

\authorrunning{The HERMES Collaboration}

\abstract{
Exclusive $\rho^0$-meson electroproduction is studied by the HERMES
experiment, using the $27.6$~GeV longitudinally polarized
electron/positron beam of HERA and
a transversely polarized hydrogen target, in the kinematic region $1.0$~GeV$^2
<Q^2<7.0$~GeV$^2$, $3.0$~GeV $<W<6.3$~GeV, and
$-t^\prime<0.4$~GeV$^2$. Using an unbinned maximum-likelihood method, 25
parameters are extracted. These determine the real
and imaginary parts of the ratios of several helicity amplitudes describing
$\rho^{0}$-meson production by a virtual photon.
The denominator of those ratios is the dominant amplitude, the
nucleon-helicity-non-flip amplitude $F_{0\frac{1}{2}0\frac{1}{2}}$,
which describes the production
of a longitudinal $\rho^{0}$-meson by a longitudinal virtual photon.
The ratios of nucleon-helicity-non-flip amplitudes are found to be in good agreement with
those from the previous HERMES analysis.
The transverse target polarization allows for the first time the extraction of
ratios of
a number of nucleon-helicity-flip amplitudes to $F_{0\frac{1}{2}0\frac{1}{2}}$.
Results obtained in a handbag approach based on generalized parton distributions 
 taking into account the contribution from pion exchange are found to be in good agreement with these ratios. Within the model, the data favor a positive sign 
for the $\pi-\rho$ transition form
factor. 
By also exploiting the longitudinal beam polarization, a total of 71 $\rho^0$ spin-density matrix elements is determined from the
extracted 25 parameters, in contrast to only 53 elements as directly determined in
earlier analyses.
}

\maketitle

\section{Introduction}
\label{sec:intro}

Exclusive electroproduction of vector mesons ($V$) on nucleons ($N$)
has been  investigated for many decades (see, for instance, Refs.~\cite{Bauer,INS}).  Originally, the 
 reaction mechanism was of primary interest, but now it has become apparent that this process  also offers the possibility to study the structure of the nucleon 
and of the 
vector meson \cite{Bauer,INS,MD}, especially at large virtuality $Q^2$ of the photon exchanged between electron and nucleon.
 In the one-photon-exchange approximation, all electroproduction observables can be expressed
in terms of the virtual-photon spin-density matrix  and  the matrix elements of the electromagnetic current between quantum states of initial and final 
hadrons. The latter matrix elements are 
called helicity amplitudes $F_{\lambda_V \lambda '_N \lambda_{\gamma} \lambda_N}$. They describe the process
\begin{equation}
\gamma^*(\lambda_{\gamma})+N(\lambda_N) \rightarrow V(\lambda_V)+N(\lambda '_N) \;, 
\label{react01}
\end{equation}
where $\gamma ^*$ denotes the virtual photon and 
the helicities of the particles are given in parentheses. 
In the present paper, the helicity amplitudes are defined  in the center-of-mass (CM) system of virtual photon and nucleon. 
The spin-density matrix of the virtual photon is well known from quantum electrodynamics and the spin-density matrix 
elements (SDMEs) of the produced vector meson, which describe its final spin states, are experimentally accessible.
This opens in principle the possibility to extract the helicity amplitudes, as it is detailed below. 

The formalism describing SDMEs of the produced vector meson was first presented in Ref.~\cite{SW} for unpolarized targets only, and expressions of SDMEs in terms 
of  helicity amplitudes were also established. The formalism was then extended to the case of polarized targets 
in Ref.~\cite{Fraas}. An alternative, general formalism for the description of the process in 
Eq.~(\ref{react01}) through SDMEs was  presented in Ref.~\cite{Diehl}. In the latter formalism, which is used throughout this paper, 
the SDMEs describing the production on an unpolarized target are denoted by 
$u^{\lambda_V \lambda'_V}_{\lambda_{\gamma}\lambda'_{\gamma}}$, those describing the production on a longitudinally polarized 
 target are denoted by $l^{\lambda_V \lambda'_V}_{\lambda_{\gamma}\lambda'_{\gamma}}$, and those describing the production on a transversely polarized target are 
denoted by $n^{\lambda_V \lambda'_V}_{\lambda_{\gamma}\lambda'_{\gamma}}$ and $s^{\lambda_V \lambda'_V}_{\lambda_{\gamma}\lambda'_{\gamma}}$. 
Here, longitudinal and transverse polarization are defined with respect to the momentum direction of the virtual photon in the CM system of the 
process in Eq.~(\ref{react01}). 

The exact expressions for SDMEs~\cite{SW,Fraas,Diehl}, which are dimensionless quantities, can be rewritten in terms of ratios of helicity amplitudes.
When fitting the experimental  angular distribution of the final-state particles, either the SDMEs or alternatively the amplitude ratios 
can be considered as independent free parameters.  
The first fit method is referred to as  the ``SDME method'' in the rest of this paper, while the second one is referred to as the ``amplitude method''.

Exclusive meson production in hard lepton-nucleon scattering was shown to offer the possibility of constraining 
 generalized parton distributions (GPDs), which provide correlated information on transverse-spatial 
and frac\-tio\-nal-longi\-tu\-dinal-momentum distributions of partons in the 
nucleon (see Refs.~\cite{MD,gpd1,VGG,GPV,GPRV,BR} and references 
therein). Vector-meson-production amplitudes contain various linear combinations of GPDs for quarks 
 of various flavors and for gluons.  
In particular, exclusive $\rho^{0}$ production on an unpolarized target is sensitive to the 
nu\-cle\-on-helicity-non-flip GPD $H$, while exclusive $\rho^{0}$ production on a transversely 
polarized target is sensitive to the nucleon-helicity-flip GPD $E$, as well. 
Through the Ji relation~\cite{gpd2}, the sum of both GPDs $H$ and $E$ is related to the parton total angular momentum. 
Access to  GPDs  relies on the factorization property of the process amplitude, 
 i.e.,  the amplitude can be written as a convolution of GPDs and vector-meson distribution amplitudes, which are both non-per\-tur\-ba\-tive quantities, 
and amplitudes of hard partonic subprocesses, which are calculable within the frameworks of perturbative quantum chromodynamics (pQCD) and  
quantum electrodynamics.

For spin-1 particles, longitudinal (transverse) polarization is assigned by convention to the states with helicity $\lambda=0$ ($\lambda = \pm 1$).
The helicity amplitudes $F_{0\frac{1}{2}0 \pm \frac{1}{2}}$ describe the transition of a longitudinally ($L$) polarized  virtual 
photon to a longitudinally polarized vector meson, $\gamma^*_L \to V_L$, and dominate at large photon virtuality $Q^2$. 
Although factorization was rigorously proven~\cite{CFS} only for these amplitudes, 
it was assumed in Refs.~\cite{golos2,golos3} 
that factorization also  holds for the amplitudes 
$F_{1\frac{1}{2}1\pm \frac{1}{2}}$ and $F_{0 \frac{1}{2}1\pm \frac{1}{2}}$, which describe the transition 
from a transversely polarized virtual photon to a 
transversely polarized meson, $\gamma^*_T \to V_T$, and a 
longitudinally polarized meson, $\gamma^*_T \to V_L$, respectively. 
The agreement found between certain calculated SDMEs and those extracted 
from  HERMES~\cite{DC-24}, ZEUS~\cite{ZEUS2}, and  H1~\cite{H1-amp} data supports this 
assumption. In general, the differential and total cross sections for $\rho^0$-meson 
production by virtual photons are reasonably well described in the GPD-based approach of 
Refs.~\cite{golos2,golos3}, not only at the high energies of the HERA collider 
experiments~\cite{ZEUS3,ZEUS4,H1-1,H1,H1-2}, but also at intermediate energies covered by the
fixed-target experiments E665~\cite{E665} and HERMES~\cite{rho-xsec}.

The real parts of the amplitude ratios in $\rho^0$ and $\phi$ meson electroproduction on the proton were first studied by the H1 experiment~\cite{H1-amp} at the HERA 
collider. In the HERMES experiment~\cite{DC-84}, $\rho^0$-meson production on unpolarized protons and deuterons was investigated. 
Both real and imaginary parts of the ratios of amplitudes without nucleon helicity flip were extracted at HERMES using a 
longitudinally polarized electron or positron beam.
The results of the analysis of $\rho^0$-meson and $\omega$-meson production on the unpolarized targets at HERMES using the SDME method were published 
in Ref.~\cite{DC-24} and Ref.~\cite{DC-95}, respectively. 
The SDMEs for the electroproduction on transversely polarized protons were published in Ref.~\cite{DC-71}. 
In this paper, the work of Ref.~\cite{DC-71} is continued. Ratios of $\rho^{0}$ helicity amplitudes with respect to the amplitude 
  $F_{0\frac{1}{2}0\frac{1}{2}}$   are extracted  from HERMES data collected with longitudinally polarized electron and positron beams scattered off transversely 
polarized protons. The amplitude ratios that require measurements with a transversely polarized target are reported for the first time in this paper.

At fixed $Q^2$ and CM energy $W$ in
the $\gamma^{*}N$ system the cross section $\text{d}\sigma/\text{d}t$, which is differential in
the Mandelstam variable $t$, contains the linear combination of squares of all
helicity amplitudes. Including $\text{d}\sigma/\text{d}t$ in an amplitude analysis of all
beam and target-polarization states would allow the extraction of the
moduli of all amplitudes and of the phase differences between them, while
the common phase would remain undetermined.

The amplitude ratios measured at HERMES, as described in this paper, will also be compared to those evaluated within the GPD-based handbag approach by Goloskokov 
and Kroll~\cite{golos2,golos3}, hereafter referred to as ``GK model".

The paper is organized as follows.
In Section~\ref{sec:formalism}, the theoretical formalism is introduced.
 Section~\ref{sec:experiment} briefly describes the experimental setup and specifies the applied data selection. The extraction procedure of the amplitude ratios 
is treated in Section~\ref{sec:amp-extrc}. The obtained results are discussed in Section~\ref{sec:results}.
Summary and conclusions are given in Section~\ref{sec:summary}. 

\section{Formalism}
\label{sec:formalism}
\subsection{Kinematics}

The process under investigation is 
\begin{equation}\label{rhoprod}
e+N\rightarrow e+\rho^0+ N,   
\end{equation}
with 
\begin{equation}\label{rhodec}
 \rho^{0} \rightarrow \pi^+ +\pi^-. 
\end{equation}
In accordance with the notation of Ref.~\cite{DC-24}, the kinematic variables of the  process under study 
are defined as follows. The four-momenta of the incident and outgoing leptons are denoted by $k$ and
$k^{\prime}$, respectively, the difference of which defines the four-momentum $q=k-k'$
of the virtual photon $\gamma^*$. 
The photon virtuality $Q^2 = -q^2$ is positive in leptoproduction. 
The squared  invariant mass of the 
photon-nucleon system is given by
\begin{equation} \label{wdef}
W^2=(p+q)^2=M^2+2\,M\,\nu-Q^2,
\end{equation}
with $M$ the nucleon mass, $p$ the four-momentum of the incident nucleon and 
\begin{equation}
\nu = \frac{p\cdot q}{M}\stackrel{lab.}{=}E-E^{\prime},
\end{equation} 
the energy transfer from the incoming lepton to the virtual photon in the target rest frame (``lab.'' frame). 
Here, $E$ ($E'$) is the energy of the incident (scattered) lepton. 

The Mandelstam variable $t$ is defined by the relation
\begin{equation}\label{eqt}
t=(q-v)^2,
\end{equation}
where $v$ is the four-momentum of the $\rho^0$ meson, equal to $p_{\pi^+}+p_{\pi^-}$, the sum of the $\pi^+$ and $\pi^-$  four-momenta. The variables $t$,
$t_0$, and $t^\prime=t-t_0$ are always negative, where  $-t_{0}$ is the minimal value of $-t$ for given values of $Q^2$, $W$, and the $\rho^0$-meson mass
$M_V$.    At small values of $-t^\prime$, the approximation $-t^\prime \approx v_T^2$ holds, where $v_T$ is the  transverse momentum of the $\rho^0$  meson
 with respect to the direction of the virtual photon in the $\gamma^{*} N$ CM system.

The variable $\epsilon$ represents the  ratio of fluxes of longitudinally and transversely polarized virtual  photons: 
\begin{eqnarray}\label{expreps}
\epsilon=\frac{1-y - \frac{Q^2}{4E^2} }{1-y+ \frac{y^2}{2} + \frac{Q^2}{4E^2}}, 
\end{eqnarray}
with $y = p\cdot q / p\cdot k   \stackrel{lab.}{=} \nu / E$.

The ``exclusivity'' of $\rho^0$ production in  the process in Eq.~(\ref{rhoprod}) is characterized by the missing energy 
\begin{equation}
\label{deltae}
\Delta E = \frac{M_{X}^{2} -M^{2}}{2 M } \stackrel{lab.} = E_V - (E_{\pi^+}+E_{\pi^-}),
\end{equation}
 where $M_X = \sqrt{ (k - k' + p - p_{\pi^+}-p_{\pi^-})^2}$ is the reconstructed invariant mass of  the undetected hadronic  system (missing mass), $E_V= \nu + 
t/(2 M )$ is the energy of the exclusively produced $\rho^0$ meson, and $(E_{\pi^+}+E_{\pi^-})$ is the sum of the energies
 of the two detected pions in the target rest frame. 

\begin{figure}[t]\centering
\includegraphics[width=8.0cm]{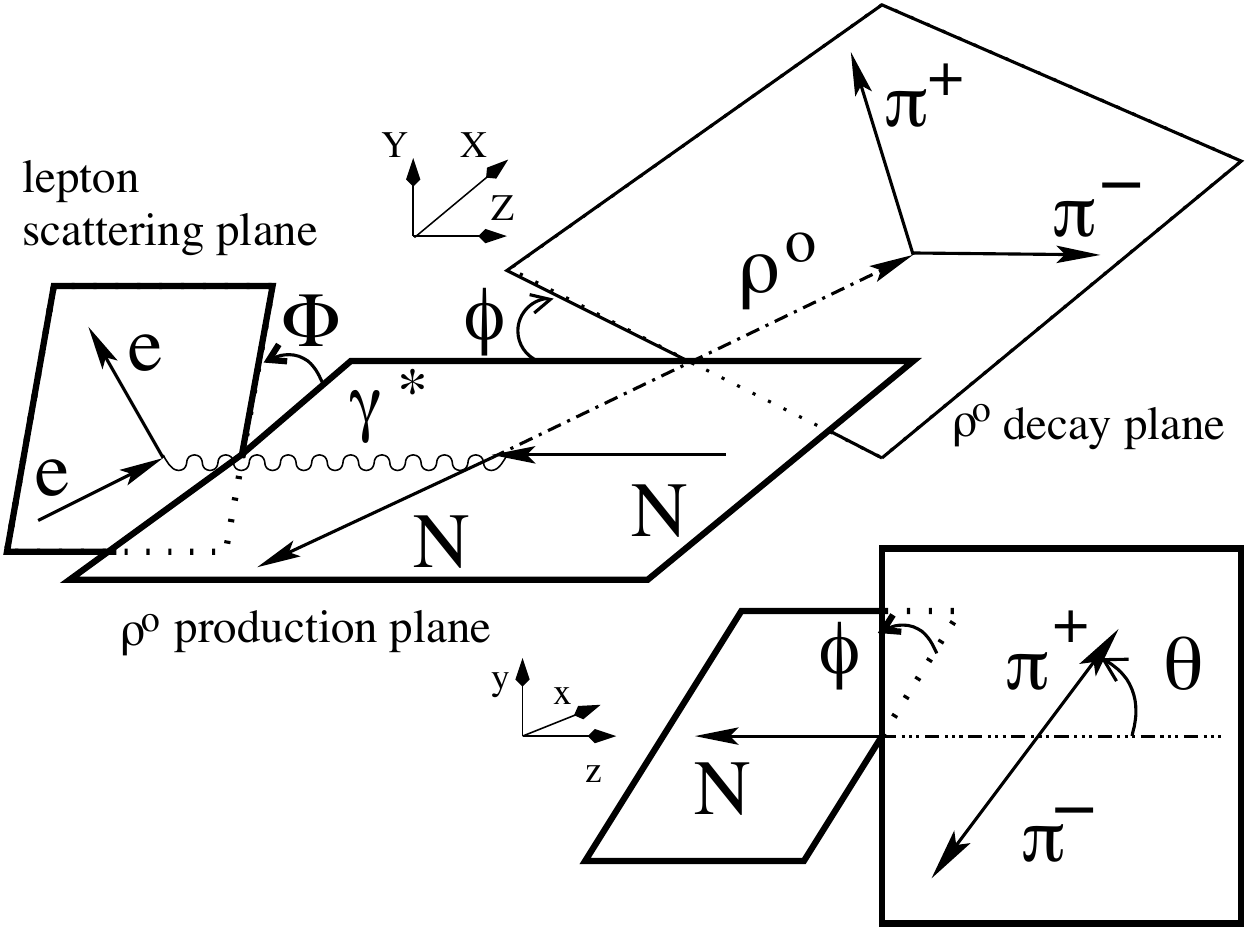}
\caption{\small
Definition of angles in the process $eN\rightarrow e\rho^{0} N \rightarrow e \pi^+ \pi^- N$. Here, $\Phi$ is
the angle between
 the $\rho^0$ production plane and the lepton scattering plane in the CM system of virtual photon and target nucleon. 
The variables $  \theta$ and $\phi$ are respectively the polar and azimuthal angles of the decay
 $\pi^+$ in the $\rho^0$-meson rest frame, with the $z$ axis being anti-parallel to the outgoing nucleon momentum. 
The $XZ$ and $xz$ planes both contain 
the $\gamma^{*}$ and $\rho^{0}$ three-momenta.} 
\label{hellen}
\end{figure}

\subsection{Definition of angles and coordinate systems}
\label{sec:def-angles}

 The  angles used for the description of the process are defined in the same way as in Ref.~\cite{DC-24}, according to Ref.~\cite{DESY2},  
 and are presented in  Fig.~\ref{hellen}. According to  Ref.~\cite{SW}, the right-handed
``hadronic CM system'' of coordinates $XYZ$ of virtual photon and target nucleon is defined such that 
the $Z$-axis  is aligned along the virtual-photon three-momentum $\vec{q}$ and the $Y$-axis is parallel to $\vec{q} \times \vec{v}$,  where $\vec{v}$ is the
$\rho^0$-meson three-momentum.
The angle $\Phi$  is the 
angle between the $\rho^0$-meson production plane ($XZ$ plane, which coincides with the 
nucleon scattering plane) and the lepton scattering plane in the CM system.
The angles $\theta$ and $\phi$ are defined in the
right-handed $xyz$ system of coordinates (see Fig.~\ref{hellen}) that represents the $\rho^0$-meson rest frame. The $y$ axis coincides with the $Y$ axis.
The angle $\theta$ is the polar angle  of the decay $\pi^+$-meson three-momentum with respect to  
 the $z$ axis,  where the latter is aligned opposite to the direction of the momentum of the outgoing nucleon. 
The azimuthal angle  of the $\pi^+$ momentum with respect to the  $\rho^0$-meson production plane 
in the CM system is denoted $\phi$. In the HERMES experiment, 
the vector $\vec{P}_T$  of the target polarization is orthogonal  to the beam direction. The 
angle  between the directions of the transverse part (with respect to the beam) of the scattered electron momentum and $\vec{P}_T$ is denoted by $\Psi$ and is 
defined in the target rest frame. 

\subsection{Natural and unnatural-parity-exchange helicity amplitudes}
\label{sec:npe-upe-ampls}

The helicity amplitudes $F_{\lambda_{V}\lambda '_N\lambda_{\gamma}\lambda _N}$ describing exclusive $\rho^0$-meson production by the virtual photon 
are here defined in the  hadronic CM system~\cite{SW}. 
 These helicity amplitudes can be expressed as scalar products of the matrix element of 
 the electromagnetic current vector $J^{\kappa}$ and the virtual-photon polarization vector $e_{\kappa}^{(\lambda_{\gamma})}$:  
\begin{eqnarray}
F_{\lambda_{V} \lambda '_{N} \lambda_{\gamma}  \lambda_{N} } = (-1)^{\lambda_{\gamma}}
\langle v \lambda_{V} p' \lambda '_{N} | J ^{\kappa} | p \lambda_{N} \rangle
e_{\kappa}^{(\lambda_{\gamma})},
 \label{jacobwick}
\end{eqnarray}
 where a summation over the Lorentz index $\kappa$ is performed. Here, $e_{\kappa}^{(\pm 1)}$ and 
 $e_{\kappa}^{(0)}$ indicate  transverse and longitudinal polarization of the virtual photon in the CM 
system, respectively:
\begin{eqnarray}
\label{transvphot}
e^{(\pm 1)}& =& (e^{(\pm 1)}_0,e^{(\pm 1)}_X,e^{(\pm 1)}_Y,e^{(\pm 1)}_Z)=(0,\mp \frac{1}{\sqrt{2}},-\frac{i}{\sqrt{2}},0)\;,
\nonumber\\
e^{(0)}&=& (e^{(0)}_0,e^{(0)}_X,e^{(0)}_Y,e^{(0)}_Z) = \frac{1}{Q}(q_Z,0,0,q_0),
 \label{longphot}   
\end{eqnarray} 
where $q_0$ and $q_Z$ are the energy and the $Z$ component of the three-momentum of the virtual photon in the CM system given by
\begin{eqnarray}
q_0=\frac{M\nu-Q^2}{W},\;\;\;q_Z=\frac{M\sqrt{\nu^2+Q^2}}{W}.
 \label{q0-qz}
\end{eqnarray}
 The ket vector $| p \lambda_{N} \rangle$ corresponds to the  initial  nucleon and the bra vector $\langle v \lambda_{V} p' \lambda '_{N} |$ 
represents the final state  consisting of a $\rho ^0$ meson and the scattered nucleon.

Any helicity amplitude $F_{\lambda_{V} \lambda '_{N} \lambda_{\gamma}  \lambda_{N} }$ can be decomposed into the 
sum of an amplitude $T_{\lambda_{V} \lambda '_{N} \lambda_{\gamma}  \lambda_{N} }$ for na\-tu\-ral-parity exchange 
(NPE) and an amplitude $U_{\lambda_{V} \lambda '_{N} \lambda_{\gamma}  \lambda_{N} }$ 
for unnatural-parity exchange (UPE)~\cite{SW,Fraas,Diehl}: 
\begin{equation}
F_{\lambda_{V} \lambda '_{N} \lambda_{\gamma}  \lambda_{N} } =
T_{\lambda_{V} \lambda '_{N} \lambda_{\gamma}  \lambda_{N} }+
U_{\lambda_{V} \lambda '_{N} \lambda_{\gamma}  \lambda_{N} },
\label{decomp}
\end{equation}
where the NPE and UPE amplitudes are defined as 
\begin{align}
\nonumber
& T_{\lambda_{V} \lambda '_{N} \lambda_{\gamma}  \lambda_{N} }   \\ 
& \quad =  \frac{1}{2}[F_{\lambda_{V} \lambda '_{N} \lambda_{\gamma}  \lambda_{N} }
+(-1)^{\lambda_{N}-\lambda'_{N}}F_{\lambda_{V} -\lambda '_{N} \lambda_{\gamma}  -\lambda_{N} }],
\label{npe-amp}\\
\nonumber
 & U_{\lambda_{V} \lambda '_{N} \lambda_{\gamma}  \lambda_{N} }  \\ 
 & \quad =  \frac{1}{2}[F_{\lambda_{V} \lambda '_{N} \lambda_{\gamma}  \lambda_{N} }
-(-1)^{\lambda_{N}-\lambda'_{N}}F_{\lambda_{V} -\lambda '_{N}\lambda_{\gamma}  -\lambda_{N} }].
\label{upe-amp}
\end{align}
These amplitudes by their definition obey the  symmetry relations 
\begin{eqnarray}
T_{\lambda_{V} \lambda '_{N}\lambda_{\gamma} \lambda_{N}} 
& = & (-1)^{\lambda '_{N}-\lambda_{N}}
T_{\lambda_{V} -\lambda '_{N}\lambda_{\gamma} -\lambda_{N}},
 \label{symmnat}\\
U_{\lambda_{V} \lambda '_{N}\lambda_{\gamma} \lambda _{N}} 
& = & -(-1)^{\lambda '_{N}-\lambda_{N}}
U_{\lambda_{V} -\lambda '_{N}\lambda_{\gamma} -\lambda _{N}}.
  \label{symmunn}
\end{eqnarray}
Equations (\ref{symmnat}) and (\ref{symmunn}) permit the introduction of the following abbreviated notations for the  amplitudes:  
\begin{eqnarray}
T^{(1)}_{\lambda_{V} \lambda_{\gamma} } & \equiv & T_{\lambda_{V} \frac{1}{2}\lambda_{\gamma} \frac{1}{2}}   = 
 T_{\lambda_{V} -\frac{1}{2}\lambda_{\gamma} -\frac{1}{2}},
  \label{tdiag}\\
U^{(1)}_{\lambda_{V} \lambda_{\gamma} } & \equiv & U_{\lambda_{V} \frac{1}{2}\lambda_{\gamma} \frac{1}{2}}  = 
-U_{\lambda_{V} -\frac{1}{2}\lambda_{\gamma} -\frac{1}{2}},
  \label{udiag}
\end{eqnarray}
which are diagonal with respect to the nucleon helicity ($\lambda _{N}=\lambda' _{N}$), and
\begin{eqnarray}
T^{(2)}_{\lambda_{V} \lambda_{\gamma} } & \equiv & T_{\lambda_{V} \frac{1}{2}\lambda_{\gamma} -\frac{1}{2}}  = 
-T_{\lambda_{V} -\frac{1}{2}\lambda_{\gamma} \frac{1}{2}}, 
  \label{tnon}\\
U^{(2)}_{\lambda_{V} \lambda_{\gamma} } & \equiv & U_{\lambda_{V} \frac{1}{2}\lambda_{\gamma} -\frac{1}{2}}  = 
U_{\lambda_{V} -\frac{1}{2}\lambda_{\gamma} \frac{1}{2}}
  \label{unon}
\end{eqnarray}   
hold for the  amplitudes with nucleon-helicity flip. 
Due to parity conservation  (see, e.g., Refs.~\cite{SW,Diehl}), 
 the amplitudes $T^{(j)}_{\lambda_{V} \lambda_{\gamma}}$ and $U^{(j)}_{\lambda_{V} \lambda_{\gamma}}$ obey the symmetry 
relations for $j=1,\;2$: 
\begin{eqnarray}
T^{(j)}_{\lambda_{V} \lambda_{\gamma} } &  =  &
(-1)^{-\lambda_{V}+\lambda_{\gamma}}\,
T^{(j)}_{-\lambda_{V} -\lambda_{\gamma} } 
  \label{symtn}, \\
U^{(j)}_{\lambda_{V} \lambda_{\gamma} }  & =   &
-(-1)^{-\lambda_{V}+\lambda_{\gamma}}\,
U^{(j)}_{-\lambda_{V} -\lambda_{\gamma}}.           
  \label{symun} 
\end{eqnarray}
This implies that there is a linear dependence between certain amplitudes. Therefore, if some property of the amplitude 
$T^{(j)}_{\lambda_{V} \lambda_{\gamma} }$ 
($U^{(j)}_{\lambda_{V} \lambda_{\gamma}}$) is established for some particular $\lambda_{V}$, $\lambda_{\gamma}$, and $j$, 
the amplitude $T^{(j)}_{-\lambda_{V} -\lambda_{\gamma} }$ ($U^{(j)}_{-\lambda_{V} -\lambda_{\gamma}}$) has the same property.

 There are three  important consequences of the symmetry relations (\ref{symtn}) and (\ref{symun})~\cite{SW,Diehl}:
\begin{enumerate}
\item The number of linearly independent NPE (UPE) amplitudes is equal to 10 (8);
\item No UPE amplitude exists for the transition $\gamma_{L} \to \rho^0_{L}$, so that in particular
$F_{0 \frac{1}{2} 0 \frac{1}{2} } = T_{0 \frac{1}{2} 0 \frac{1}{2} } \equiv T^{(1)}_{00}$;
\item For unpolarized targets there is no interference between NPE and UPE amplitudes~\cite{SW,Diehl}. 
\end{enumerate}

 At small values of $t$, in Regge phenomenology~\cite{IW,KS} the exchange of a single natural-parity reggeon, i.e., with parity  $P=(-1)^J$, such as a pomeron or 
secondary reggeons $\rho$, 
$f_2$, $a_2$, ..., contributes only to the NPE amplitudes. The exchange of a single unnatural-parity reggeon, i.e., with parity 
  $P=-(-1)^J$, such as $\pi$, $a_1$, $b_1$,...,  contributes only to the UPE amplitudes~\cite{CSM}. 

\subsection{Asymptotic behavior of amplitudes at small $|t'|$}

Considering only the behavior of a helicity amplitude at $-t' \to 0$, 
its magnitude relative to other amplitudes can be investigated by using the following parametrization  
\begin{eqnarray}
 F_{\lambda_V \lambda'_N \lambda_{\gamma} \lambda_N}=\sum_{k=0}^{\infty} c^{\lambda_V \lambda'_N \lambda_{\gamma} \lambda_N}_{k} \Bigl 
(\frac{\sqrt{-t'}}{M_h}\Bigr)^{s+2k}, 
\label{asytpr}
\end{eqnarray}
where $s=|(\lambda_V-\lambda'_N)-(\lambda_{\gamma}-\lambda_N)|$ and $M_{h}$ represents the typical hadronic mass of the order of $1$~GeV. 
If $c^{\lambda_V \lambda'_N \lambda_{\gamma} \lambda_N}_{0}$ vanishes, then 
the power series with respect to $t'$ starts with a term $\propto (\sqrt{-t'}/M_h)^{s+2k}$ with $k\geq 1$.
The asymptotic relations for $T_{\lambda_{V} \lambda '_{N}\lambda_{\gamma} \lambda _{N}}$ and $U_{\lambda_{V} \lambda 
'_{N}\lambda_{\gamma} \lambda _{N}}$, and hence for $T^{(1)}_{\lambda_{V} \lambda_{\gamma} }$, $T^{(2)}_{\lambda_{V} \lambda_{\gamma} }$,
$U^{(1)}_{\lambda_{V} \lambda_{\gamma} }$, and $U^{(2)}_{\lambda_{V} \lambda_{\gamma} }$, follow from 
Eq.~(\ref{asytpr}) and  Eqs.~(\ref{npe-amp}) and (\ref{upe-amp}).

From the asymptotic behavior, the NPE amplitudes $T^{(1)}_{00}$, $T^{(1)}_{11}$, $T^{(2)}_{01}$, $T^{(2)}_{10}$ and
the UPE amplitudes $U^{(1)}_{11}$, $U^{(2)}_{01}$, $U^{(2)}_{10}$  are proportional to $(-t')^0$ at $|t'|\to 0$. Therefore, the amplitude ratios
\begin{eqnarray}
t^{(1)}_{11},\;t^{(2)}_{01},\;t^{(2)}_{10},\;u^{(1)}_{11},\;u^{(2)}_{01},\;u^{(2)}_{10}
\label{prop-t0}
\end{eqnarray}
defined by the relations
\begin{eqnarray}
t^{(j)}_{\lambda_V \lambda_{\gamma}} & = & T^{(j)}_{\lambda_V \lambda_{\gamma}}/T^{(1)}_{00},
\label{def-tjk}\\
u^{(j)}_{\lambda_V \lambda_{\gamma}} & = & U^{(j)}_{\lambda_V \lambda_{\gamma}}/T^{(1)}_{00}
\label{def-ujk}
\end{eqnarray}
for $j=1,\;2$ can  be non-zero for $|t'|\rightarrow 0$. These ratios are expected to attain their largest values at small $-t'$.

Similarly, one can conclude from Eq.~(\ref{asytpr}) that the amplitude ratios proportional to $\sqrt{-t'}/M_h$ at small $-t'$ are 
\begin{eqnarray}
t^{(2)}_{00},\;t^{(2)}_{ 1 1},\;t^{(1)}_{0  1},\;t^{(1)}_{ 10},\;u^{(2)}_{ 11},\; u^{(2)}_{ 1 -1},\;u^{(1)}_{0 1},\;u^{(1)}_{ 10}.
\label{prop-t1}
\end{eqnarray}
 However, if for a UPE amplitude appearing in Eq.~(\ref{prop-t1}) the pion exchange in the $t$ channel is significant, the typical 
scale for $t'$ is about
 $m_{\pi}^2$. Therefore, this amplitude can be of the order of the dominant amplitude $T^{(1)}_{00}$ at $-t' \sim m_{\pi}^2$.  
 Hence some amplitude ratios from Eq.~(\ref{prop-t1}) can be of the same order of magnitude as those in Eq.~(\ref{prop-t0}) at $-t' 
\sim m_{\pi}^2$. 

The smallest amplitudes at $|t'| \to 0$ are the double spin-flip amplitudes $T^{(1)}_{1 -1}$ and $U^{(1)}_{1 -1}$ and 
the amplitude ratios
\begin{eqnarray}
t^{(1)}_{ 1 -1},\;u^{(1)}_{1 -1}
\label{prop-t2}
\end{eqnarray}  
are proportional to $-t'$.

\subsection{Spin-density matrix of the virtual photon}

 The  spin-density matrix of the virtual photon, normalized such that the flux of transversely polarized photons is equal to unity, 
embodies the unpolarized ($U$) and 
 polarized ($L$) matrices~\cite{SW}:
\begin{equation} \label{matr}
\varrho^{U+L}_{\lambda_{\gamma} \lambda '_{\gamma }} =
\varrho_{\lambda_{\gamma}
\lambda '_{\gamma }}^{U} +
P_B \; \varrho_{\lambda_{\gamma} \lambda '_{\gamma}}^{L},
\end{equation}
with 
\begin{align}
& \varrho^{U}_{\lambda_{\gamma} \lambda '_{\gamma}}(\epsilon, \Phi) \nonumber \\
& = \frac{1}{2} \! \left(\!\! \begin{array}{ccc}
1 & \sqrt{\epsilon(1+\epsilon)} e^{-i\Phi} & -\epsilon e^{-2i\Phi} \\
\sqrt{\epsilon(1+\epsilon)} e^{i\Phi} & 2 \epsilon & - \sqrt{\epsilon(1+\epsilon)}
e^{-i\Phi} \\
- \epsilon  e^{2 i\Phi} & - \sqrt{\epsilon(1+\epsilon)} e^{i\Phi} & 1 \\
\end{array} \! \right) \!,  \label{matr-unpol} \\
& \varrho^{L}_{\lambda_{\gamma} \lambda '_{\gamma}}(\epsilon, \Phi) \nonumber \\
& = \frac{\sqrt{1-\epsilon}}{2} \left( \begin{array}{ccc}
\sqrt{1+\epsilon}  & \sqrt{\epsilon} e^{-i\Phi} & 0 \\
\sqrt{\epsilon} e^{i\Phi} &  0  & \sqrt{\epsilon} e^{-i\Phi} \\
0 & \sqrt{\epsilon} e^{i\Phi} & -\sqrt{1+\epsilon}  \\
\end{array} \right ), \label{matr-lpol}
\end{align}
and where $P_B$ is the longitudinal polarization of the beam and $\epsilon$ is defined by Eq.~(\ref{expreps}). 
In the above formulas, the spin-density matrices of the virtual photon are defined in the hadronic CM system.

\subsection{Spin-density matrix of the initial nucleon}

The angle between the three-momenta of the initial electron and the virtual photon in the target rest frame  
$\theta_{\gamma}$ can be calculated as 
\begin{eqnarray}
\cos \theta_{\gamma} & = & \frac{\nu+Q^2/(2E)}{\sqrt{\nu^2+Q^2}},
\label{csthetgam}\\
\sin \theta_{\gamma} & = & \frac{Q \sqrt{1-y-Q^2/(4E^2)}}{\sqrt{\nu^2+Q^2}}. 
\label{snthetgam}
\end{eqnarray}
The ``hadronic rest system'', in which the initial proton is at rest, has the $\hat{X}$, $\hat{Y}$, and $\hat{Z}$ 
axes parallel to the $X$, $Y$, and $Z$ axes of the hadronic CM system, defined in Section~\ref{sec:def-angles} and shown in Fig.~\ref{hellen}.
The components of the target polarization vector $\vec{P}$ in the hadronic rest system are 
\begin{eqnarray}
\hat{P}_X & = & P_T(\cos \Phi \cos \Psi \cos \theta_{\gamma}-\sin \Phi  \sin \Psi),
\label{px}\\
\hat{P}_Y & = & P_T(\cos \Phi  \sin \Psi +\sin \Phi \cos \Psi \cos \theta_{\gamma}),
\label{py}\\  
\hat{P}_Z & = & -P_T\cos \Psi  \sin \theta_{\gamma}. 
\label{pz}
\end{eqnarray}
The spin-density matrix of the initial nucleon in the helicity representation can be written in this system as
\begin{equation}
\tau_{\lambda_N \lambda'_N}  = \frac{1}{2}
 \begin{pmatrix}
 1-\hat{P}_Z& \hat{P}_X+i\hat{P}_Y\\  
 \hat{P}_X-i\hat{P}_Y& 1+\hat{P}_Z
 \end{pmatrix}.
\label{taun}
\end{equation}
Since the nucleon spin is anti-parallel to the $\hat{Z}$ axis for $\lambda_N=\frac{1}{2}$ and parallel to the $\hat{Z}$ axis for 
$\lambda_N=-\frac{1}{2}$,  Eq.~(\ref{taun}) for the spin-density matrix of the initial nucleon does not coincide with the standard
formula $(I+\vec{P}\vec{\sigma})/2$ for the quantization axis aligned along the $\hat{Z}$ axis. Here, 
$\vec{\sigma}=(\sigma_x,\sigma_y,\sigma_z)$ are the Pauli matrices and $\vec{P}$ is the 
polarization vector in the target rest frame. 
The hadronic CM system can be obtained from the hadronic rest system by a boost along the virtual-photon three-momentum, 
which is antiparallel to the proton three-momentum. Since the value of the proton helicity is invariant under this boost
the spin-density matrix is also boost invariant, hence it  is given by Eq.~(\ref{taun}) 
in the hadronic CM system. 

\subsection{Spin-density matrix of the $\rho^{0}$ meson}

The spin-density matrix $\rho_{\lambda_{V} \lambda '_{V}}$
of the produced $\rho^{0}$ meson  is
related through the von~Neumann formula 
to those of the virtual photon, 
$\varrho^{U+L}_{\lambda_{\gamma} \lambda '_{\gamma }}$, and the nucleon, $\tau_{\lambda _N \lambda' _N}$: 
\begin{eqnarray} 
\rho_{\lambda_{V} \lambda '_{V}} =   
\sum
\frac{F_{\lambda_{V}\mu _N\lambda_{\gamma}\lambda _N}\;
 \varrho^{U+L}_{\lambda_{\gamma} \lambda '_{\gamma }}\;\tau_{\lambda _N \lambda' _N}
  F_{\lambda '_{V} \mu _N\lambda '_{\gamma}\lambda' _N}^{*}}{2\mathcal{N}},  
\label{neumann}
 \end{eqnarray}
where the sum runs over $\lambda_{\gamma}$, $\lambda '_{\gamma}$, $\lambda_N$, $ \lambda '_N$, and  $\mu_N$.
The normalization factor is given by
\begin{eqnarray}
\mathcal{N} =   \mathcal{N}_T + \epsilon \mathcal{N}_L,
\label{ntotal}
\end{eqnarray}
with
\begin{eqnarray}
\nonumber
\mathcal{N}_{T}  = \sum_{j=1,2} & &
 ( |T^{(j)}_{11}|^2+|T^{(j)}_{01}|^2+|T^{(j)}_{-11}|^2 \\
&&  + |U^{(j)}_{11} |^2+|U^{(j)}_{01}|^2+|U^{(j)}_{-11}|^2 ),
\label{norm-t}\\
\mathcal{N}_L  = \sum_{j=1,2} & & ( |T^{(j)}_{00}|^2+2|T^{(j)}_{10}|^2+2|U^{(j)}_{10}|^2 ).
\label{norm-l}
\end{eqnarray}
Equation (\ref{norm-l}) is obtained by using the symmetry 
relations (\ref{symtn}) and (\ref{symun}).

\subsection{SDMEs in the Diehl representation}\label{sec-2.8}

The spin-density matrix elements calculated below are defined in accordance with Ref.~\cite{Diehl} as
\begin{eqnarray}
\nonumber
u^{\lambda_V \lambda'_V}_{\lambda_{\gamma}\lambda '_{\gamma}}=\frac{1}{\mathcal{N}}
\sum_{\sigma=\pm \frac{1}{2}} \Bigl [ T_{\lambda_{V} \sigma \lambda_{\gamma} \frac{1}{2}} 
\Bigl ( T_{\lambda '_{V} \sigma \lambda '_{\gamma} \frac{1}{2}} \Bigr )^* \\
+U_{\lambda_{V} \sigma \lambda_{\gamma} \frac{1}{2}} 
\Bigl ( U_{\lambda '_{V} \sigma \lambda '_{\gamma} \frac{1}{2}} \Bigr )^*
\Bigr ],
\label{diehl-u}\\
\nonumber
l^{\lambda_V \lambda'_V}_{\lambda_{\gamma}\lambda '_{\gamma}}=\frac{1}{\mathcal{N}}
\sum_{\sigma=\pm \frac{1}{2}} \Bigl [ T_{\lambda_{V} \sigma \lambda_{\gamma} \frac{1}{2}}
\Bigl ( U_{\lambda '_{V} \sigma \lambda '_{\gamma} \frac{1}{2}} \Bigr )^* \\
+U_{\lambda_{V} \sigma \lambda_{\gamma} \frac{1}{2}}
\Bigl ( T_{\lambda '_{V} \sigma \lambda '_{\gamma} \frac{1}{2}} \Bigr )^*
\Bigr ],
\label{diehl-l}\\
\nonumber
s^{\lambda_V \lambda'_V}_{\lambda_{\gamma}\lambda '_{\gamma}}=\frac{1}{\mathcal{N}}
\sum_{\sigma=\pm \frac{1}{2}} \Bigl [ T_{\lambda_{V} \sigma \lambda_{\gamma} \frac{1}{2}}
\Bigl ( U_{\lambda '_{V} \sigma \lambda '_{\gamma} -\frac{1}{2}} \Bigr )^* \\
+U_{\lambda_{V} \sigma \lambda_{\gamma} \frac{1}{2}}
\Bigl ( T_{\lambda '_{V} \sigma \lambda '_{\gamma} -\frac{1}{2}} \Bigr )^*
\Bigr ],
\label{diehl-s}\\
\nonumber
n^{\lambda_V \lambda'_V}_{\lambda_{\gamma}\lambda '_{\gamma}}=\frac{1}{\mathcal{N}}
\sum_{\sigma=\pm \frac{1}{2}} \Bigl [ T_{\lambda_{V} \sigma \lambda_{\gamma} \frac{1}{2}}
\Bigl ( T_{\lambda '_{V} \sigma \lambda '_{\gamma} -\frac{1}{2}} \Bigr )^* \\
+U_{\lambda_{V} \sigma \lambda_{\gamma} \frac{1}{2}}
\Bigl ( U_{\lambda '_{V} \sigma \lambda '_{\gamma} -\frac{1}{2}} \Bigr )^*
\Bigr ].
\label{diehl-n}
\end{eqnarray}
Here, the NPE and UPE helicity amplitudes are defined in Eqs.~(\ref{npe-amp}) and (\ref{upe-amp}), and $\mathcal{N}$ denotes the normalization factor
given by Eqs.~(\ref{ntotal}-\ref{norm-l}).

\subsection{Angular distribution of decay pions}
\label{sec-3.5}

The angular distribution $\mathcal{W}(\Phi, \Psi,\theta, \phi)$ of the pions from the $\rho^0$-meson decay 
in Eq.~(\ref{rhodec}), which are produced in the process (\ref{rhoprod}), is related to the spin-density matrix
$\rho_{\lambda_{V} \lambda '_{V}}$ through 
\begin{eqnarray}
 \mathcal{W}(\Phi, \Psi,\theta, \phi)=\!\!\!\!\sum_{\lambda_{V} \lambda '_{V}}\rho_{\lambda_{V} \lambda '_{V}}Y_{1\lambda_{V}}(\theta, 
\phi)
Y^*_{1\lambda'_{V}}(\theta, \phi),
\label{wtot}
\end{eqnarray}
where $Y_{1\lambda_{V}}$ is the spherical function. The phases of
$Y_{1\lambda_{V}}$ are chosen as in Ref. \cite{Diehl}: 
\begin{eqnarray}
\nonumber
Y_{1 \pm 1}(\theta, \phi)=\mp\sqrt{\frac{3}{8 \pi}} \sin \theta e^{ \pm i \phi},\;\;
Y_{1 0}(\theta, \phi)=\sqrt{\frac{3}{4 \pi}} \cos \theta .
\end{eqnarray} 
These phases determine the phases of the extracted helicity amplitudes.
 Equation~(\ref{wtot}) shows explicitly the dependence of the angular distribution $\mathcal{W}$ on $\theta$ and $\phi$, while the 
dependences 
on $\Phi$ and $\Psi$ are hidden in the kinematic dependences of the spin-density matrix $\rho_{\lambda_{V}\lambda '_{V}} $. 
The angular distribution 
depends on the kinematic variables $W$, $Q^2$, and $-t'$ through the dependence of the helicity amplitudes on these variables  in 
Eq.~(\ref{neumann}).  
For simplicity of notation, these dependences are omitted throughout the paper. 

Since $\tau_{\lambda_N \lambda'_N }$ depends linearly on
the nucleon transverse polarization $P_T$, and $\varrho_{\lambda_{\gamma}\lambda'_{\gamma}}$ is a linear  
function of the beam polarization, the formula for the angular distribution
contains four terms:
\begin{eqnarray}
\mathcal{W}=\mathcal{W}_1+\mathcal{W}_2P_B+\mathcal{W}_3P_T+\mathcal{W}_4P_BP_T.
\label{w1-4}
\end{eqnarray}
Note that the angular dependent functions $\mathcal{W}_m$ for any $m$ in Eq.~(\ref{w1-4}) are themselves independent of $P_B$ and $P_T$.

 It can easily be shown that the angular distribution $\mathcal{W}(\Phi, \Psi,\theta, \phi)$ cannot be negative for any set of values 
of the complex amplitudes 
$F_{\lambda_{V}\mu _N\lambda_{\gamma}\lambda _N}$, even for unphysical ones. This property is of great importance for the fit procedure. 
It is worthwhile to note that using the SDME method one faces the problem of a possible negativity of 
$\mathcal{W}(\Phi, \Psi,\theta, \phi)$ for some angles when SDMEs assume unphysical values. As it is unknown in which 
region in the multi-dimensional space of SDMEs 
 $\mathcal{W}(\Phi, \Psi,\theta, \phi)$ is not negative, serious problems may appear when applying the maximum-likelihood method. 
Hence the amplitude method is in that respect more reliable than the SDME method. 

Altogether, Eqs. (\ref{wtot}), (\ref{neumann}-\ref{norm-l}), (\ref{matr}-\ref{matr-lpol}), and (\ref{csthetgam}-\ref{taun}),  
with the substitutions $T_{00}^{(1)}\rightarrow 1$ and for all other amplitudes 
$T_{\lambda_{V}\lambda_{\gamma}}^{(j)}\rightarrow t_{\lambda_{V}\lambda_{\gamma}}^{(j)}$ 
and $U_{\lambda_{V}\lambda_{\gamma}}^{(j)}\rightarrow u_{\lambda_{V}\lambda_{\gamma}}^{(j)}$, constitute a 
basis for the amplitude method, in which the extracted quantities are the helicity-amplitude ratios.  

\section{Experiment and data selection}
\label{sec:experiment}
\subsection{Experiment}

 A detailed description of the HERMES experiment can be found in Ref.~\cite{herspec}. The data analyzed in this paper were collected 
between the years 2002 and 2005. 
 A longitudinally polarized positron or electron beam of $27.6$~GeV was scattered from a pure gaseous, transversely polarized hydrogen 
target internal to the HERA 
lepton storage ring. The helicity of the beam was typically reversed every two months. 
The beam polarization was continuously measured by two Compton polarimeters~\cite{BAR,BECK}. The average value of the 
beam polarization for the events used in the analysis is about $\pm 0.3$, with a relative uncertainty of 2\%. 
The target polarization  was reversed every 60~s to 180~s~\cite{HER-POL}.
The measured mean value of the target polarization is $\langle|P_{T}|\rangle=0.72 \pm 0.06$~\cite{Ami-th,Jeron-th}. 

 The HERMES setup included a forward spectrometer~\cite{herspec}, in which the scattered lepton and the produced hadrons were detected 
within an angular acceptance of
  $\pm$170~mrad horizontally and $\pm(40 - 140)$~mrad vertically. The tracking system had a momentum resolution of about $1.5\%$ and an 
angular resolution 
 of about $1$~mrad. Lepton identification was accomplished using a transition-radiation detector, a  preshower scintillator counter, 
and an electromagnetic 
 calorimeter.  The particle-identification system included also a dual-radiator ring-imaging Che\-ren\-kov  detector~\cite{RICH} to 
identify hadrons. 
Combining the responses of the detectors in a likelihood method leads to an average lepton-identification efficiency of 98\%, 
with a hadron contamination of less than 1\%. 

\subsection{Event selection}

 The event sample used in this analysis is almost the same as that used in Ref.~\cite{DC-71}. The most important improvement is the 
application 
of a new tracking algorithm, which is based on a Kalman filter~\cite{Kal}.
For the present analysis, the data are required to fulfill the following criteria:
\begin{enumerate}
\item The longitudinal beam polarization is restricted to the interval $15\% < | P_B | < 80\%$.
\item Events with exactly two oppositely charged hadrons and one lepton with the same charge as the beam lepton are selected. 
All tracks are required to originate from the same vertex. 
\item The scattered lepton has to have an energy larger than 3.5 GeV in order to not introduce effects from varying trigger thresholds.
\item The two-hadron invariant mass is required to lie a\-round the $\rho^{0}$ mass, i.e., it is required to  obey 
$0.6$~GeV $<M(\pi^+\pi^-)<1.0$~GeV.
 \item The photon virtuality is required to obey 1 GeV$^2<Q^2<  7$ GeV$^2$. The lower limit is a minimum requirement for the 
application of pQCD, while the upper one delimits a well defined kinematic phase space.   
\item The $t'$ variable is restricted to $-t' \leq 0.4$ GeV$^2$ in order to reduce non-exclusive background of the reaction under study.
\item The invariant mass $W$ is required to obey 3~GeV $< W \leq 6.3$~GeV. The 
requirement $W>3$ GeV is imposed in order to be outside of the resonance region. 
The upper constraint delimits a well defined kinematic phase space. 
\item For exclusive $\rho^{0}$-meson production,  $\Delta E$ as defined in Eq.~(\ref{deltae}) must vanish. 
In the present analysis, taking into consideration the 
 spectrometer resolution, the missing energy has to be in the region  $-1.0$ GeV $ < \Delta E < 0.6$ GeV. This region is referred to as 
``exclusive region'' in the following.
\end{enumerate}

After application of all these constraints, the data sample contains 8741 events. These data are referred to in the 
following as data in the ``entire kinematic region''.
 The applied requirements do not fully suppress contributions from background processes. The exclusive sample contains contributions 
from non-resonant $\pi^+ \pi^-$ pair production, which
 is of the order of $1-2\%$ \cite{Ami-th}, and from semi-inclusive deep-inelastic scattering (SIDIS) events. The presented results are 
not corrected for the former process, 
while a correction is applied for SIDIS background.  
 The uncertainty of the correction for background from SIDIS events is considered to be one of the main contributions to the total 
systematic uncertainty.

\section{Extraction of amplitude ratios}  
\label{sec:amp-extrc}

\begin{table*}[hbt!]
 \renewcommand{\arraystretch}{1.2}
\begin{center}
\begin{tabular}{|c|c|c|c|c|}
\hline
Cell limits&$\langle W \rangle$, GeV& $\langle Q^2 \rangle$, GeV$^2$&$ \langle -t^\prime \rangle$, GeV$^2$&$f_{bg}$\\
\hline
$1.0$ GeV$^2 <Q^2 <1.4$ GeV$^2$&4.70 &1.19 &0.128&0.065 \\
\hline
$1.4$ GeV$^2 <Q^2 <2.0$ GeV$^2$&4.75 &1.67 &0.128&0.073 \\
\hline
$2.0$ GeV$^2 <Q^2 <7.0$ GeV$^2$&4.80 &3.06 &0.136&0.122 \\
\hline
$0.00$ GeV$^2 <-t' <0.05$ GeV$^2$&4.75 &1.89 &0.023&0.064 \\
\hline
$0.05$ GeV$^2 <-t' <0.10$ GeV$^2$&4.75 &1.92 &0.074&0.085 \\
\hline
$0.10$ GeV$^2 <-t' <0.20$ GeV$^2$&4.71 &1.94 &0.145&0.108 \\
\hline
$0.20$ GeV$^2 <-t' <0.40$ GeV$^2$&4.72 &2.00 &0.281&0.147 \\
\hline
\end{tabular}
\\[2pt]
\caption{ \label{binning} {Mean values for the kinematic variables $W$, $Q^2$, $-t'$,
and $f_{bg}$ under the exclusive peak in each of the ($Q^2$, $-t'$) cells.}}
\end{center}
\end{table*}

\subsection{Fit of the angular distribution}
\label{sec:fit}

A maximum-likelihood method is used to fit the angular distribution of the $\rho^0$-meson decay pions and the scattered lepton. 
The probability to measure $\rho^0$ decay pions in the small angular region $\text{d} \Omega=\text{d} \Phi \text{d} \Psi \sin \theta \text{d} \theta \text{d} \phi$ 
is proportional to $\mathcal{W}(\Phi, \Psi,\theta, \phi) \mathcal{E}(\Phi, \Psi,\theta, \phi)\text{d} \Omega$. Here, the detector efficiency is denoted by 
$\mathcal{E}(\Phi, \Psi,\theta, \phi)$. It includes geometric detector acceptance, particle-detection and par\-ti\-cle-identification efficiencies as well as 
track-re\-con\-struc\-tion efficiency. In order to become a probability, this expression needs to be normalized to unity: 
\begin{eqnarray}
\text{d}w(\mathcal{R})=\frac{\mathcal{W}(\mathcal{R},P_B,P_T,\Phi, \Psi,\theta, \phi) \mathcal{E}(\Phi, \Psi,\theta, \phi)\text{d} \Omega}
{\int \mathcal{W}(\mathcal{R},P_B,P_T,\Phi, \Psi,\theta, \phi) \mathcal{E}(\Phi, \Psi,\theta, \phi)\text{d} \Omega}.\nonumber\\
\;
\label{prob}
\end{eqnarray}
Here, we have changed the notation of the angular distribution to $\mathcal{W}(\mathcal{R},P_B,P_T,\Phi, \Psi,\theta, \phi)$, 
as $\mathcal{W}$ depends on the set $\mathcal{R}$ of all amplitude ratios (due to Eqs.~(\ref{wtot}), (\ref{neumann}) and 
(\ref{ntotal}-\ref{norm-l})) as well as on the beam and target polarization. 
All factors that are independent of the set of amplitude ratios can be omitted in the expression of 
the likelihood function. 
The likelihood function is evaluated in $3 \times 4$ ($Q^2,-t'$) cells. Within each of the  
12 cells, the detector efficiency is considered to be independent of $Q^2$ and $-t'$. 
The lower and upper boundaries of the cells and the mean values of 
the variables $W$, $Q^2$, and $-t'$ are presented in Table~\ref{binning}. 
Within each of these cells, the logarithm 
of the likelihood function can be written as the sum over all experimental events in this cell ($i=1, 2, ..., I$):
\begin{eqnarray}
\ln\mathcal{L}(\mathcal{R})=\sum_{i=1}^{I}\ln[\mathcal{W}(\mathcal{R},P_{Bi},P_{Ti},\Phi_i, \Psi_i,\theta_i, \phi_i)/N_i(\mathcal{R})].\nonumber\\
\; 
\label{loglike}
\end{eqnarray}  
Here, $N_i$ is the normalization factor for the $i$-th event in the corresponding $(Q^2,-t')$ cell, with 
\begin{eqnarray}
N_i(\mathcal{R})=\int \mathcal{W}(\mathcal{R},P_{Bi},P_{Ti},\Phi, \Psi,\theta, \phi) \mathcal{E}(\Phi, \Psi,\theta, \phi)\text{d} \Omega.\nonumber\\
\;
\label{statnorm}
\end{eqnarray}
Using the expression for $\mathcal{W}$ from Eq.~(\ref{w1-4}), the 
normalization factor can be represented as
\begin{align}
  N_i(\mathcal{R})  & =   \mathcal{K}_1(\mathcal{R})  \nonumber\\
 + &\, \mathcal{K}_2(\mathcal{R})P_{Bi}+\mathcal{K}_3(\mathcal{R})P_{Ti}+\mathcal{K}_4(\mathcal{R})P_{Bi}P_{Ti}, \label{norm1-4}
\end{align}
where the functions $\mathcal{K}_m$ ($m=1,\;2,\;3,\;4$) are given by the integrals 
\begin{eqnarray} 
\mathcal{K}_m(\mathcal{R})=\int \mathcal{W}_m(\mathcal{R},\Phi, \Psi,\theta, \phi) \mathcal{E}(\Phi, \Psi,\theta, \phi)\text{d} \Omega.
\label{k1-4}
\end{eqnarray} 
 In order to evaluate the integrals in Eqs.~(\ref{k1-4}), a dedicated PYTHIA Monte Carlo (MC)~\cite{pythia} simulation was used to generate exclusively produced 
$\rho^0$  events 
with a uniform angular distribution, hereafter referred to as ``uniform exclusive $\rho^0$ MC''~\cite{Ami-th,DC-24}. 
The total number of generated events is by a factor of one hundred larger than that of the experimental data. The same event selection 
requirements are applied to the reconstructed events from the simulation as to the events from experimental data.  
The integrals in Eqs.~(\ref{k1-4}) can be approximated using the MC as follows:  
\begin{eqnarray}
\mathcal{K}_m(\mathcal{R}) \approx 
\frac{C}{L}\sum_{l=1}^L \mathcal{W}_m(\mathcal{R},\Phi_l, \Psi_l,\theta_l, \phi_l).
\label{mc-k1-4}
\end{eqnarray}
Here, $L$ is the total number of reconstructed MC events and $C$ at $L \to \infty$ is a constant equal to $2 (2 \pi)^3$, which is the total volume
for the variables $\cos \theta$, $\Phi$, $\Psi$, and $\phi$. 
Since this constant does not 
depend on the free parameters, the know\-ledge of its value is unimportant for the maximum-likelihood method and it is  
set to unity below. In order to avoid computer calculations with very large values when building up the sum in
Eqs.~(\ref{mc-k1-4}), the right-hand side is divided by $L$. 
The value of $\mathcal{K}_m(\mathcal{R})$ is evaluated separately in each of the $3 \times 4$ ($Q^2,-t'$) cells. 
 According to their experimental occurrence, the various settings of target and  beam polarizations are assigned as $(P_{B},P_{T})=(+1,+1),$ $(+1,-1),$ 
$(-1,+1)$, and $(-1,-1)$ to the generated MC events. 
Thus four independent equations for $N_i$ can be obtained in each ($Q^2,-t'$) cell 
and the four functions $\mathcal{K}_m(\mathcal{R})$ can then be determined in each of the corresponding cells.
The value of $\mathcal{K}_m(\mathcal{R})$ is  also evaluated separately for electron and positron data. 
The finally maximized logarithm of the likelihood function is the sum of the logarithms of the likelihood function from Eq.~(\ref{loglike})  
over all cells for both electron and positron data.

\subsection{Background corrections}

One of the main sources of background contamination to exclusive $\rho^0$-meson electroproduction  
in deep-inelastic scattering originates from SIDIS. 
A PYTHIA MC using GEANT3~\cite{geant} to simulate the HERMES apparatus and tuned to the kinematics of the HERMES experiment~\cite{th-pati}, hereafter referred to as ``SIDIS MC'', 
is used for the estimation of this background contribution. The same kinematic and geometrical requirements are
imposed on both simulated and real data samples. The normalization of the MC data to the experimental data 
is performed in the region $2$~GeV $< \Delta E < 20 $~GeV (see Fig.~3 in Ref. \cite{DC-71}), 
and the number of background events in the exclusive region 
is estimated. The fraction of SIDIS 
background $f_{bg}$ as estimated from the SIDIS MC is shown in the fifth column in Table~\ref{binning}.

 It is assumed that the angular distribution of the SIDIS
background events is reasonably well reproduced by the SIDIS MC simulation. 
The fit of the angular distribution of the SIDIS MC events under the exclusive peak for each ($Q^2,-t'$) cell is performed using 
Eq.~(\ref{loglike}) in which the substitutions $\mathcal{W} \to \mathcal{W}_{bg}$, $N_i \to N^{bg}_i$,  
and $\mathcal{R} \to \mathcal{S}$ must be performed, and the sum runs over all background MC events. 
The set of free parameters $\mathcal{S}$ represents the complete set of 15 ``unpolarized''
$u^{\lambda_V \lambda'_V}_{\lambda_{\gamma}\lambda'_{\gamma}}$ SDMEs 
describing the background. 
The normalization factor for the background, $N_{i}^{bg}$, is determined in an analogous way as done for the signal events, 
but in the present analysis the background angular distribution is considered to be independent of the beam and target 
polarizations, i.e.,  $P_B$ and $P_T$ are set to zero.
After the fit of the SIDIS MC events, the angular distribution of the 
background is considered to be fixed, hence $\mathcal{W}_{bg}(\mathcal{S},\Phi_i, \Psi_i,\theta_i, \phi_i)$, $N^{bg}_i$, and $f_{bg}$ 
do not contain any free parameters. 

The total probability $\text{d}w_{tot}$ to measure final-state particles from the reaction in Eqs.~(\ref{rhoprod}) and~(\ref{rhodec})  or 
from SIDIS background in the small angular region $\text{d} \Omega$ is given by 
\begin{eqnarray}
\text{d}w_{tot}(\mathcal{R})=(1-f_{bg})\text{d}w+f_{bg}\text{d}w_{bg},
\label{totprob}
\end{eqnarray} 
where $\text{d}w$ is given by Eq.~(\ref{prob}) and $\text{d}w_{bg}$ is its analogue for the background process. 
In a similar manner as done in Eq.~(\ref{loglike}), the  
logarithm of the likelihood function $\mathcal{L}_{tot}$, which  
takes into account the background events, is given by 
\begin{align}
\ln \mathcal{L}_{tot}(\mathcal{R})= \sum_{i=1}^I \ln\Bigl[ &(1-f_{bg})
\frac{\mathcal{W}(\mathcal{R},P_{Bi},P_{Ti},\Phi_i, \Psi_i,\theta_i, \phi_i)}{N_i} \nonumber \\
& +f_{bg}\frac{\mathcal{W}_{bg}(\mathcal{S},\Phi_i, \Psi_i,\theta_i, \phi_i)}{N^{bg}_i} \Bigr].
\label{logliketot}
\end{align}  
The logarithms of the likelihood function are again calculated separately in each $(Q^{2},-t')$ cell for both 
 electron and positron data, and the finally maximized logarithm of the likelihood function is the sum of the logarithms of the likelihood function from 
Eq.~(\ref{logliketot}) over all cells for both electron and positron data. 
As a result, the amplitude ratios in the entire kinematic region are obtained. 

\subsection{Choice of free parameters}
\label{sec:choice}

As explained in Section~\ref{sec:intro}, the angular distribution of the detected particles depends on
the amplitude ratios. 
The total number of linearly independent helicity-amplitude ratios defined by Eqs.~(\ref{def-tjk}-\ref{def-ujk}) is 17,
which means that 34 real functions of $Q^2$, $-t'$, and $W$ determine all SDMEs and angular distributions. As established in Ref.~\cite{DC-84},
the large amplitudes at $-t' \leq 0.4$ GeV$^2$ and $Q^2 \geq 1$ GeV$^2$ are 
$T^{(1)}_{00}$, $T^{(1)}_{11}$, $U^{(1)}_{11}$, and $T^{(1)}_{01}$. For the ratios of large amplitudes, the parameterization of
the $Q^2$ and $-t'$ dependences is chosen as in Ref.~\cite{DC-84}. For the small amplitude ratios, only the $-t'$ dependence following from angular momentum 
conservation (see for instance Ref.~\cite{Diehl}) is taken into account, while averaging over the kinematic range in $Q^2$.
If all other amplitudes are 
expected to be significantly smaller than the large amplitudes, a possibility to extract the small amplitudes exists only if they are 
multiplied by large amplitudes. 
This means that they contribute linearly to the angular distribution.

The easiest way to interpret the extractability of the various helicity-amplitude ratios is through their contribution to the SDMEs, as detailed in Ref.~\cite{Diehl}.
For the transversely polarized target, the SDMEs $n^{\lambda_V \lambda'_V}_{\lambda_{\gamma}\lambda'_{\gamma}}$ and
$s^{\lambda_V \lambda'_V}_{\lambda_{\gamma}\lambda'_{\gamma}}$ contribute~\cite{Diehl}, 
while the contribution of the SDMEs $l^{\lambda_V \lambda'_{V}}_{\lambda_{\gamma}\lambda'_{\gamma}}$ 
is neglected in this analysis, since the latter are multiplied by the longitudinal component of the target polarization $|\hat{P}_Z|$.
Indeed, $|\hat{P}_Z|$ is proportional to $\sin \theta _{\gamma}$ (see Eq.~(\ref{pz})), 
which in turn is proportional to $Q/\nu$ (according to Eq.~(\ref{snthetgam})). 
At HERMES kinematics $Q$ is much smaller than $\nu$, with  $Q/\nu$ of the order of 0.1. 
The helicity-amplitude ratios $t^{(2)}_{\lambda_V \lambda_{\gamma}}$ and $u^{(2)}_{\lambda_V \lambda_{\gamma}}$ can
be extracted from data collected with a transversely polarized target, as they contribute linearly to the SDMEs
$n^{\lambda_V \lambda'_V}_{\lambda_{\gamma}\lambda'_{\gamma}}$ and $s^{\lambda_V \lambda'_V}_{\lambda_{\gamma}\lambda'_{\gamma}}$, respectively. 
Contributions of squares of moduli of the helicity-amplitude ratios $u^{(1)}_{01}$, $u^{(1)}_{10}$, $u^{(1)}_{1-1}$ 
to $u^{\lambda_V \lambda'_{V}}_{\lambda_{\gamma}\lambda'_{\gamma}}$ are much smaller than the contribution of $|u^{(1)}_{11}|^2$ 
according to the hierarchy of amplitudes established  in Refs.~\cite{DC-24} and \cite{DC-84}. 
The small heli\-city amplitude ratios $u^{(1)}_{\lambda_V \lambda_{\gamma}}$ with $\lambda_V \neq \lambda_{\gamma}$ contribute linearly to
the SDMEs $l^{\lambda_V \lambda'_V}_{\lambda_{\gamma}\lambda'_{\gamma}}$. The latter are multiplied by the small factor $\hat{P}_Z \sqrt{1-\epsilon}$ and cannot be 
extracted from the angular distributions of final-state pions. 
Therefore the helicity-amplitude ratios $u^{(1)}_{\lambda_V \lambda_{\gamma}}$, with the exception of $u^{(1)}_{11}$, are set equal to zero in the 
fit.

For an unpolarized target, only the
SDMEs $u^{\lambda_V \lambda'_V}_{\lambda_{\gamma}\lambda'_{\gamma}}$ contribute to the angular distribution. As follows from the previous 
 analysis at HERMES~\cite{DC-84} the ratios $t^{(1)}_{\lambda_V \lambda_{\gamma}}$ and $|u^{(1)}_{11}|^2+|u^{(2)}_{11}|^2$ can be reliably extracted 
from data collected with an unpolarized target. 
The value of  $|u^{(1)}_{11}|$ can be extracted from the unpolarized data, since the numerators of some SDMEs
$u^{\lambda_V \lambda'_V}_{\lambda_{\gamma}\lambda'_{\gamma}}$ contain  $|u^{(1)}_{11}|^2+|u^{(2)}_{11}|^2$.
However, the phase $\delta_u$ of  $u^{(1)}_{11}$ cannot be obtained reliably given the limited statistics in the
present analysis.
Another function that cannot be reliably extracted from the present data is ${\rm{Im}}\{t^{(1)}_{11}\}$. The reason is that
it contributes mainly to the imaginary parts of the
SDMEs $u^{\lambda_V \lambda'_V}_{\lambda_{\gamma}\lambda'_{\gamma}}$, which are multiplied by the small factor $P_B\sqrt{1-\epsilon}$. 
This factor is smaller than $0.15$, since $\epsilon$ is about $0.8$ and the mean value of $|P_B|$ is about $0.3$.

If ${\rm{Im}}\{t^{(1)}_{11}\}$ and the phase of $u^{(1)}_{11}$ are considered as free parameters, the fit to the angular
distribution becomes unstable, leading to several local minima in the fit.
 In order to avoid such instabilities, the function ${\rm{Im}}\{t^{(1)}_{11}\}$ is taken from the previous analysis at HERMES~\cite{DC-84}, 
where the number of events and the value of $|P_B|$ were larger than those in the present analysis, the fit was stable, and the 
minimum was unique. In order to take into account the $Q^2$ dependence of ${\rm{Im}}\{t^{(1)}_{11}\}$,
the parametrization 

\begin{equation}
{\rm{Im}}\{t^{(1)}_{11}\}=bQ, 
\label{bQpara}
\end{equation}
with $b$ taken from Ref.~\cite{DC-84} is used.
 The phase $\delta_u$ is fixed from the data collected with a longitudinally polarized hydrogen target~\cite{HERL1,HERL2}.
A detailed discussion of this problem is given in Appendix~\ref{ap_all}.

\begin{table*}[hbt!]
 \renewcommand{\arraystretch}{1.2}
\begin{center}
\begin{tabular}{|c|c|c|c|}
\hline
Parametrization& value of parameter&statistical uncertainty &total uncertainty\\
\hline
${\rm{Re}}\{t^{(1)}_{11}\}=b_1/Q$&$b_1=1.145$ GeV&0.033 GeV&0.081 GeV\\
\hline
$|u^{(1)}_{11}|=b_2$&$b_2=0.333$ &0.016 &0.088  \\
\hline
${\rm{Re}}\{u^{(2)}_{11}\}=b_3$&$b_3=-0.074$ &0.036 &0.054  \\
\hline
${\rm{Im}}\{u^{(2)}_{11}\}=b_4$&$b_4=0.080$ &0.022 &0.037  \\
\hline
$\xi=b_5$&$b_5=-0.055$ &0.027 &0.029  \\
\hline
$\zeta=b_6$&$b_6=-0.013$ &0.033 &0.044  \\
\hline
${\rm{Im}}\{t^{(2)}_{00}\}=b_7$&$b_7=0.040$ &0.025 &0.030  \\
\hline
${\rm{Re}}\{t^{(1)}_{01}\}=b_8\sqrt{-t'}$&$b_8=0.471$ GeV$^{-1}$ &0.033 GeV$^{-1}$ &0.075 GeV$^{-1}$ \\
\hline
${\rm{Im}}\{t^{(1)}_{01}\}=b_9\frac{\sqrt{-t'}}{Q}$&$b_9=0.307 $ &0.148  &0.354   \\
\hline
${\rm{Re}}\{t^{(2)}_{01}\}=b_{10}$&$b_{10}=-0.074$ &0.060 &0.080  \\
\hline
${\rm{Im}}\{t^{(2)}_{01}\}=b_{11}$&$b_{11}=-0.067$ &0.026 &0.036  \\
\hline
${\rm{Re}}\{u^{(2)}_{01}\}=b_{12}$&$b_{12}=0.032$ &0.060 &0.072  \\
\hline
${\rm{Im}}\{u^{(2)}_{01}\}=b_{13}$&$b_{13}=0.030$ &0.026 &0.033  \\
\hline
${\rm{Re}}\{t^{(1)}_{10}\}=b_{14}\sqrt{-t'}$&$b_{14}=-0.025$ GeV$^{-1}$ &0.034 GeV$^{-1}$ &0.063 GeV$^{-1}$ \\
\hline
${\rm{Im}}\{t^{(1)}_{10}\}=b_{15}\sqrt{-t'}$&$b_{15}=0.080$ GeV$^{-1}$ &0.063 GeV$^{-1}$ &0.118 GeV$^{-1}$  \\
\hline
${\rm{Re}}\{t^{(2)}_{10}\}=b_{16}$&$b_{16}=-0.038$ &0.026 &0.030  \\
\hline
${\rm{Im}}\{t^{(2)}_{10}\}=b_{17}$&$b_{17}=0.012$ &0.018 &0.019  \\
\hline
${\rm{Re}}\{u^{(2)}_{10}\}=b_{18}$&$b_{18}=-0.023$ &0.030 &0.039  \\
\hline
${\rm{Im}}\{u^{(2)}_{10}\}=b_{19}$&$b_{19}=-0.045$ &0.018 &0.026  \\
\hline
${\rm{Re}}\{t^{(1)}_{1-1}\}=b_{20}\frac{(-t')}{Q}$&$b_{20}=-0.008$ GeV$^{-1}$ &0.096 GeV$^{-1}$ &0.212 GeV$^{-1}$ \\
\hline
${\rm{Im}}\{t^{(1)}_{1-1}\}=b_{21}\frac{(-t')}{Q}$&$b_{21}=-0.577$ GeV$^{-1}$ &0.196 GeV$^{-1}$ &0.428 GeV$^{-1}$ \\
\hline
${\rm{Re}}\{t^{(2)}_{1-1}\}=b_{22}$&$b_{22}=0.059$ &0.036 &0.047  \\
\hline
${\rm{Im}}\{t^{(2)}_{1-1}\}=b_{23}$&$b_{23}=0.020$ &0.022 &0.026  \\
\hline
${\rm{Re}}\{u^{(2)}_{1-1}\}=b_{24}$&$b_{24}=-0.047$ &0.035 &0.039  \\
\hline
${\rm{Im}}\{u^{(2)}_{1-1}\}=b_{25}$&$b_{25}=0.007$ &0.022 &0.029  \\
\hline
\end{tabular}
\\[2pt]
\caption{ 
  \label{param25} 
  Parametrization of the helicity-amplitude ratios and parameter values extracted from the fit. 
  The combinations of the helicity-amplitude ratios $\xi$ and $\zeta$ are defined in Eq.~(\ref{25-par}).
  An additional scale uncertainty of 8\% originating from the uncertainty on the target polarization is present 
  for the ratios $t^{(2)}_{\lambda_V \lambda_{\gamma}}$, $u^{(2)}_{\lambda_V\lambda_{\gamma}}$, $\xi$ and $\zeta$, but not shown. 
  An extra scale uncertainty of 2\% originating from the uncertainty on the beam polarization is present
  for the ratios 
  ${\rm{Im}}\{t^{(1)}_{\lambda_V\lambda_{\gamma}}\}$, ${\rm{Re}}\{t^{(2)}_{\lambda_V\lambda_{\gamma}}\}$, 
  ${\rm{Re}}\{u^{(2)}_{\lambda_V\lambda_{\gamma}}\}$, and $\zeta$, but also not shown.The correlations between the 25 parameters 
  are listed in Table~\ref{corrmatr} in Appendix~\ref{sec-8.4}.
}
\end{center}
\end{table*}

It is shown in Appendix~\ref{sec-8.3} that
\begin{eqnarray}
\kappa=\frac{1}{2}{\rm{Re}}\{\frac{{t}^{(2)}_{11}}{{t}^{(1)}_{11}}+t^{(2)}_{00}\}  
\label{kappa}
\end{eqnarray}
does not contribute linearly to the angular distribution and hence is set to zero.

In the fit, 25 parameters $b_{i}$ (see Table~\ref{param25}) are extracted, which determine the following 25 real functions:
\begin{align} 
\nonumber
&{\rm{Re}}\{t^{(1)}_{11}\}, \;{\rm{Re}}\{t^{(1)}_{10}\}, \;{\rm{Im}}\{t^{(1)}_{10}\}, \;{\rm{Re}}\{t^{(1)}_{1-1}\}, \; {\rm{Im}}\{t^{(1)}_{1-1}\}, \nonumber\\
&{\rm{Re}}\{t^{(1)}_{01}\}, \;{\rm{Im}}\{t^{(1)}_{01}\}, \;|u^{(1)}_{11}|,
\;\xi={\rm{Im}}\{\frac{{t}^{(2)}_{11}}{{t}^{(1)}_{11}}\}, \nonumber\\
&{\rm{Im}}\{t^{(2)}_{00}\},\;\zeta=\frac{1}{2}{\rm{Re}}\{t^{(2)}_{00} -\frac{{t}^{(2)}_{11}}{{t}^{(1)}_{11}}\},
\; {\rm{Re}}\{t^{(2)}_{10}\}, \;{\rm{Im}}\{t^{(2)}_{10}\},\nonumber\\
&{\rm{Re}}\{t^{(2)}_{1-1}\}, \;{\rm{Im}}\{t^{(2)}_{1-1}\},
{\rm{Re}}\{t^{(2)}_{01}\}, \;{\rm{Im}}\{t^{(2)}_{01}\},\nonumber\\
&{\rm{Re}}\{u^{(2)}_{11}\}, \;{\rm{Im}}\{u^{(2)}_{11}\},
{\rm{Re}}\{u^{(2)}_{10}\}, \;{\rm{Im}}\{u^{(2)}_{10}\},\nonumber\\
&{\rm{Re}}\{u^{(2)}_{1-1}\}, \;{\rm{Im}}\{u^{(2)}_{1-1}\},
\;{\rm{Re}}\{u^{(2)}_{01}\}, \;{\rm{Im}}\{u^{(2)}_{01}\}. \label{25-par}
\end{align}
 Note that ${\rm{Im}}\{\frac{{t}^{(2)}_{11}}{{t}^{(1)}_{11}}\}$ is used rather than ${\rm{Im}}\{t^{(2)}_{11}\}$, since the 
 latter is not independent of $\kappa$ and the inclusion of $\kappa$ in the fit leads to a divergence of the fit.
 Table~\ref{param25} also shows the resulting parameters with their uncertainties. The correlations between the 25 parameters are listed in Table~\ref{corrmatr} 
in Appendix~\ref{sec-8.4}.

\subsection{Systematic uncertainties}

In this subsection, the sources of systematic uncertainties and their effect on the extracted amplitude ratios are 
discussed. All systematic uncertainties except the one due to the uncertainty on the target and beam polarization measurements are added in quadrature 
to calculate the total systematic uncertainty. 
The statistical uncertainty and
the total systematic uncertainty are added in quadrature to form the total uncertainty.

\subsubsection{Systematic uncertainties due to beam and target polarization uncertainties}

The measured mean value of the target polarization is 
 $\langle|P_{T}|\rangle=0.72 \pm 0.06$~\cite{Ami-th,Jeron-th}, i.e., the fractional uncertainty of the target polarization amounts to $0.08$. The ratios 
$t^{(2)}_{\lambda_V \lambda_{\gamma}}$ and
 $u^{(2)}_{\lambda_V \lambda_{\gamma}}$ have a corresponding scale uncertainty of 8\%, since through 
their linear contribution  to the  ``transverse'' SDMEs 
$n^{\lambda_V\lambda'_V}_{\lambda_{\gamma}\lambda'_{\gamma}}$ and $s^{\lambda_V\lambda'_V}_{\lambda_{\gamma}\lambda'_{\gamma}}$, 
they are multiplied by $\langle|P_{T}|\rangle$.   
It was checked that the amplitude 
ratios $t^{(1)}_{11}$, $t^{(1)}_{10}$, $t^{(1)}_{1-1}$, $t^{(1)}_{01}$, and $|u^{(1)}_{11}|$, which 
can be extracted from data taken with an unpolarized target (see Ref.~\cite{DC-84}), are effectively    
insensitive to the uncertainty on the target polarization.

The fractional uncertainty on the beam polarization amounts to $2\%$~\cite{Sob}. This results in an additional scale uncertainty 
on ${\rm{Im}}\{u^{\lambda_V\lambda'_V}_{\lambda_{\gamma}\lambda'_{\gamma}}\}$, 
 ${\rm{Re}}\{n^{\lambda_V\lambda'_V}_{\lambda_{\gamma}\lambda'_{\gamma}}\}$, and ${\rm{Re}}\{s^{\lambda_V\lambda'_V}_{\lambda_{\gamma}\lambda'_{\gamma}}\}$  of 
$2\%$, since these SDMEs enter the expression of the angular distribution of final-state particles multiplied by the 
beam polarization~\cite{Diehl}. From the expression of SDMEs in terms of helicity-amplitude ratios, it follows that  
there is an additional scale uncertainty of $2\%$ for 
 ${\rm{Im}}\{t^{(1)}_{\lambda_V\lambda_{\gamma}}\}$, ${\rm{Re}}\{t^{(2)}_{\lambda_V\lambda_{\gamma}}\}$, and 
${\rm{Re}}\{u^{(2)}_{\lambda_V\lambda_{\gamma}}\}$, while  the influence of the uncertainty on the beam 
polarization can be neglected for 
${\rm{Re}}\{t^{(1)}_{\lambda_V\lambda_{\gamma}}\}$, ${\rm{Im}}\{t^{(2)}_{\lambda_V\lambda_{\gamma}}\}$, and 
${\rm{Im}}\{u^{(2)}_{\lambda_V\lambda_{\gamma}}\}$. 
The scale uncertainty arising from the uncertainty on the beam and target polarizations is not shown in the figures but quoted separately.

\subsubsection{Systematic uncertainty due to the extraction method}
\label{sec-sys_ex}

In order to estimate the uncertainty due to the extraction method, yet another MC data sample was produced using  
 a uniformly distributed angular distribution for exclusive $\rho^0$ production, which in contrast to the unweighted one used above for the normalization 
procedure, is weighted in order to mimic experimental data. 
It is obtained with the accept/reject method based on the experimental angular distribution and making use of the relevant parameters extracted from data. 
Using values of $\pm0.30$ for the beam polarization and $\pm 0.72$ for the target polarization, it is analyzed in a way similar 
to the experimental data. The MC sample is divided into 20 independent sets such that each set contains the same number of exclusive events as the experimental data. 
In order to evaluate the systematic uncertainty, the difference between the output value of the $j$-th 
amplitude ratio in the $k$-th set $r_{j,k}^{(out)}$ and the input value of the same amplitude ratio $r_j^{(in)}$ is compared  with the statistical 
uncertainty $\delta r_{j,k}^{(out)}$ of the output amplitude ratio in the $k$-th set. 
Averaging over the twenty sets, the relation
\begin{eqnarray}
    (\Delta r_{j}^{meth})^2=\frac{\sum_{k=1}^{K_0}\Bigl[(r_{j,k}^{(out)}-r_{j}^{(in)})^2-(\delta r_{j,k}^{(out)})^2\Bigr]}{K_0}  
\label{syst-meth2}
\end{eqnarray}
with $K_0 =20$ is used to calculate the systematic uncertainty due to the extraction method. 
If the sum in Eq.~(\ref{syst-meth2}) is positive, then the obtained value of  $\Delta r^{meth}_{j}$ is set as systematic uncertainty;  
otherwise the systematic uncertainty is set to zero. 

 \subsubsection{Systematic uncertainty due to the background contribution}

The helicity-amplitude ratios are extracted from the experimental data once taking into account the background contribution 
(see Eq.~(\ref{logliketot})) and once neglecting this contribution (see Eq.~(\ref{loglike})). 
The systematic uncertainty from the background contribution of each  amplitude ratio is computed as the modulus of the difference of the amplitude
ratios obtained for these two cases. This  conservative
approach is used, since the background correction is estimated from MC data instead of experimental data.  
The SDMEs 
$n^{\lambda_V \lambda'_V}_{\lambda_{\gamma} \lambda'_{\gamma}}$ and $s^{\lambda_V \lambda'_V}_{\lambda_{\gamma} \lambda'_{\gamma}}$ 
 are, as shown in Ref.~\cite{DC-71}, much less sensitive 
to the background contribution than the SDME $u^{\lambda_V \lambda'_V}_{\lambda_{\gamma} \lambda'_{\gamma}}$, 
since they enter the formula for the angular distribution multiplied by the target polarization. 
As the amplitude ratios $t^{(2)}_{\lambda'_N \lambda_N}$ and $u^{(2)}_{\lambda'_N \lambda_N}$
contribute linearly to these SDMEs, they are expected to be less sensitive to the background contribution 
than the amplitudes relevant for scattering off an unpolarized target. It was checked that this is indeed the case.
The small influence of the background correction to the nucleon-helicity-flip amplitude ratios
 $t^{(2)}_{\lambda'_N \lambda_N}$ and $u^{(2)}_{\lambda'_N \lambda_N}$ can be explained by the statistical correlations between these amplitude ratios and 
$t^{(1)}_{\lambda'_N \lambda_N}$, $|u^{(1)}_{11}|$. 

\subsubsection{Systematic uncertainty due to the omission of inaccessible amplitude ratios}
\label{sec-nuar}

Another source of systematic uncertainty originates from setting $u^{(1)}_{10},\;u^{(1)}_{01},\;u^{(1)}_{1-1}$
and $\kappa$, given by Eq.~(\ref{kappa}), equal to zero in the fit. 
In order to estimate this systematic uncertainty, the following procedure is applied. Since  $|T^{(2)}_{00}/T^{(1)}_{00}|$ and 
$|T^{(2)}_{11}/T^{(1)}_{11}|$ are proportional to $\sqrt{-t'}/(2M)$,  
the unmeasured parameter $\kappa$ is estimated as 
\begin{eqnarray}
\kappa=\pm\sqrt{-t'}/(2M).
\label{est-rat-28}
\end{eqnarray}
The calculations are performed for both signs in Eq.~(\ref{est-rat-28}) and the corresponding systematic uncertainties are averaged in quadrature (see  
Eq.~(\ref{syst-no})).  

For one-pion exchange, the ratios of the UPE amplitudes $U^{(1)}_{10}/U^{(1)}_{11}$, $U^{(1)}_{01}/U^{(1)}_{11}$, and $U^{(1)}_{1-1}/U^{(1)}_{11}$ are 
known. Supposing that these ratios have approximately the same value for the full amplitudes, one obtains
\begin{eqnarray}
u^{(1)}_{10} & \approx & -\frac{\sqrt{2}Qv_T}{Q^2+m^2_{\rho}}|u^{(1)}_{11}| e^{ i \delta_u},
\label{ope-1}\\
u^{(1)}_{01} & \approx & -\frac{\sqrt{2}m_{\rho}v_T}{Q^2+m^2_{\rho}}|u^{(1)}_{11}| e^{i \delta_u},
\label{ope-2}\\
u^{(1)}_{1-1} & \approx &  -\frac{v^2_T}{Q^2+m^2_{\rho}}|u^{(1)}_{11}| e^{i \delta_u}.
\label{ope-3}
\end{eqnarray}
For the calculation of the systematic uncertainty, the value $|u^{(1)}_{11}|$ is obtained from the fit in the present analysis, while the phase shift
$\delta_{u}=-39.2$ degrees, which is taken from the results of the measurement of the 
longitudinal double-spin asymmetry in exclusive $\rho^0$-meson electroproduction from Refs.~\cite{DC-24,HERL1,HERL2},  
corresponds to the value $A_1^{\rho}=0.24$ (see appendix~\ref{ap_all}). The
systematic uncertainty due to the omission of all inaccessible amplitude ratios is estimated from the relation
\begin{eqnarray}
(\Delta r_j^{inac})^2=[(r_{j}-r_{j}^{(in+)})^2+(r_{j}-r_{j}^{(in-)})^2]/2,
\label{syst-no} 
\end{eqnarray}
where the values of $r_{j}$ are
those obtained in the 25-pa\-ra\-me\-ter fit when all the amplitudes not extracted are set to zero.
The ratios $r_{j}^{(in+)}$ and $r_{j}^{(in-)}$ 
denote the extracted $j$-th amplitude ratio obtained 
in the fit in which the amplitude ratios $u^{(1)}_{10}$, $u^{(1)}_{01}$, and $u^{(1)}_{1-1}$ are calculated using Eqs.~(\ref{ope-1}-\ref{ope-3}), 
while $\kappa$ is taken according to Eq.~(\ref{est-rat-28}), once with positive sign and once with negative sign, respectively.

\subsubsection{Systematic uncertainty due to the experimental uncertainty of $A_1^{\rho}$}
\label{sec-eua1}

The uncertainty on the asymmetry $A_1^{\rho}$, as explained in appendix~\ref{ap_all}, leads to a range of  $\delta_{u}$  
between $-26.2$ degrees and $-51.1$ degrees, corresponding to $A_1^{\rho}=0.24+0.14$ and $A_1^{\rho}=0.24-0.14$. 

In order to estimate the systematic uncertainty of the obtained amplitude ratios due to the experimental uncertainty of $A_1^{\rho}$,  the 25-parameter fits with 
$\delta_{u}=-26.2$ degrees and
$\delta_{u}=-51.1$ degrees are performed in addition to the fit with $\delta_{u}=-39.2$ 
degrees. Two sets of the amplitude ratios $r_{j}^{(+)}$ and  $r_{j}^{(-)}$ are obtained, correspondingly. 
 The systematic uncertainty is calculated as 
\begin{eqnarray}
(\Delta r_j^{asym})^2=[(r_{j}-r_{j}^{(+)})^2+(r_{j}-r_{j}^{(-)})^2]/2.
\label{syst-delta_u}
\end{eqnarray}  
The systematic uncertainty due to the experimental uncertainty on $A_1^{\rho}$   
is less than the statistical uncertainty. It was found that the most important ratios $t^{(1)}_{\lambda_V \lambda_{\gamma}}$ 
and $|u^{(1)}_{11}|$ are almost insensitive to the value of $A_1^{\rho}$. 

\subsubsection{Systematic uncertainty due to the experimental uncertainty of ${\rm{Im}}\{t^{(1)}_{11}\}$}
\label{sec-imt11}

The parameter $b$, which enters the parametrization of ${\rm{Im}}\{t^{(1)}_{11}\}$ 
as given in~Eq. (\ref{bQpara}), is equal to $(0.340 \pm 0.025)$ GeV$^{-1}$~\cite{DC-84}.
Its uncertainty is another source of systematic uncertainty.
The basic fit  performed with $b=0.340$  GeV$^{-1}$ gives the amplitude ratios $r_{j}$, while the values
$(b=0.340 \pm 0.025)$ GeV$^{-1}$ correspond to the ratios $r_{j}^{(im+)}$ and $r_{j}^{(im-)}$.
The systematic uncertainty of the helicity-amplitude ratio $r_{j}$ is calculated as
\begin{eqnarray}
(\Delta r_j^{{\rm{Im}}\{t\}})^2=[(r_{j}-r_{j}^{(im+)})^2+(r_{j}-r_{j}^{(im-)})^2]/2.
\label{syst-imt11}  
\end{eqnarray}

\begin{figure*}\centering
\includegraphics[width=12.0cm]{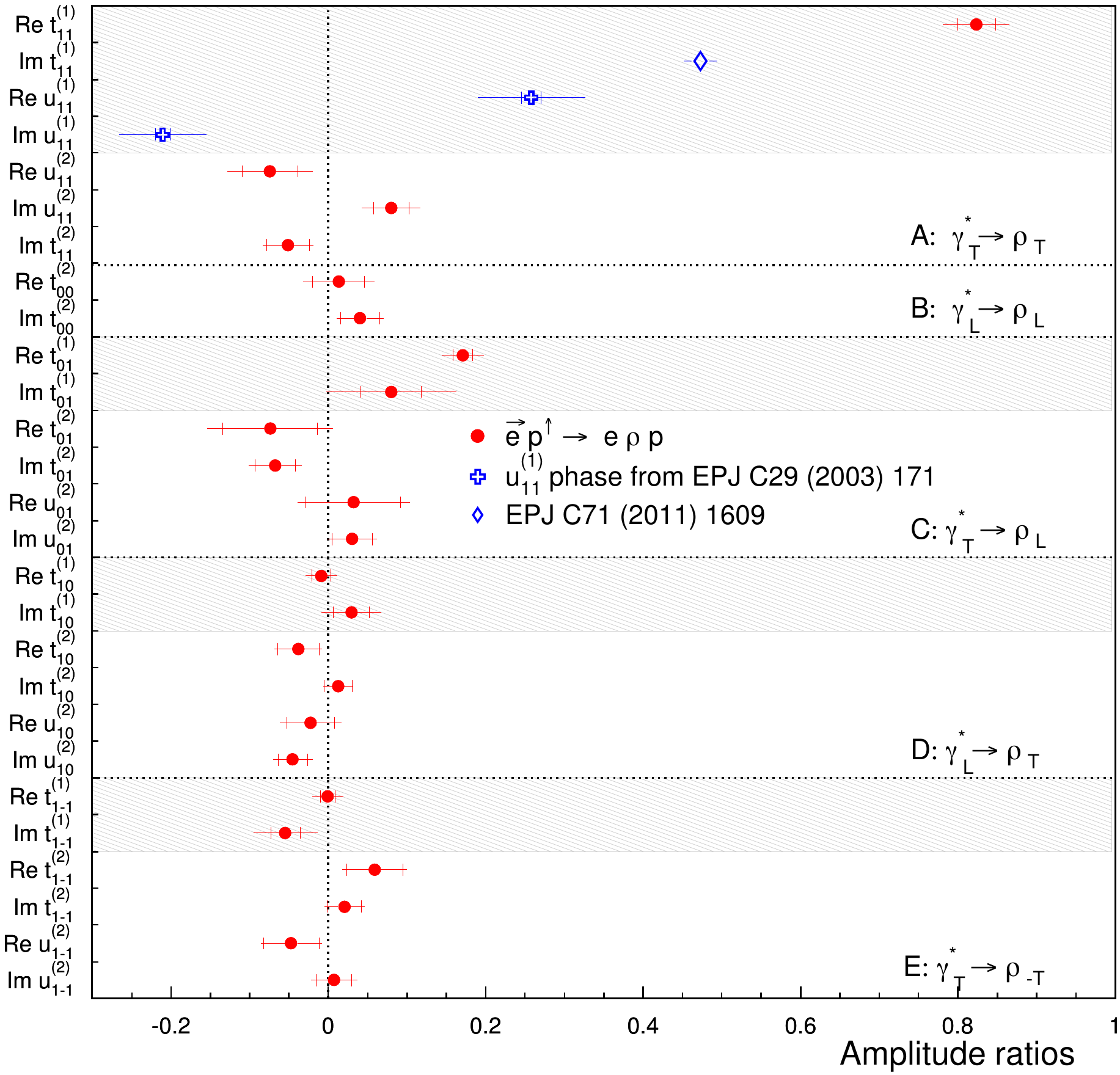}
\caption{\small Helicity-amplitude ratios obtained from the 25-parameter fit characterized by $\langle W \rangle=4.73$ GeV, $\langle Q^2\rangle=1.93$ GeV$^2$, $\langle-t'\rangle=0.132$ GeV$^2$, as explained in the text.
While the phase of $u^{(1)}_{11}$ is fixed according to the results of Refs.~\cite{DC-84,HERL1,HERL2}, its modulus is fit so that the two crosses represent the results of fitting one free parameter.
The value of ${\rm{Im}}\{t^{(1)}_{11}\}$ (open diamond) represents the result of Ref.~\cite{DC-84}; the error bar 
shows the total uncertainty. 
For all other points, the inner error bars represent the statistical  uncertainty, while the outer
ones represent statistical and systematic uncertainties added in quadrature. 
An additional scale uncertainty of 8\% originating from the uncertainty on the target polarization is present 
for the ratios $t^{(2)}_{\lambda_V \lambda_{\gamma}}$, $u^{(2)}_{\lambda_V\lambda_{\gamma}}$, but not shown. 
An extra scale uncertainty of 2\% originating from the uncertainty on the beam polarization is present
for the ratios 
${\rm{Im}}\{t^{(1)}_{\lambda_V\lambda_{\gamma}}\}$, ${\rm{Re}}\{t^{(2)}_{\lambda_V\lambda_{\gamma}}\}$ and 
${\rm{Re}}\{u^{(2)}_{\lambda_V\lambda_{\gamma}}\}$, but also not shown.
The shaded area corresponds to results that were also obtained in Ref.~\cite{DC-84}, while all other points are obtained for the first time.
The helicity-amplitude ratios are ordered according to the SDME classes proposed in Refs.~\cite{DC-24,Ami-th}. 
}
\label{rel-ent-bin-25}
\end{figure*}

\section{Results on the amplitude ratios}
\label{sec:results}

\subsection{Discussion of results on the amplitude ratios} 
\label{sec:resul_disc} 

The result obtained from the 25-parameter fit is presented in Fig.~\ref{rel-ent-bin-25}. The results for the large amplitudes are calculated at 
$-t'=0.132$~GeV$^2$ and $Q^2=1.93$~GeV$^2$, while integrating $W$ over the entire kinematic region.
The results for the small amplitudes are calculated at  $-t'=0.132$~GeV$^2$, while integrating $Q^2$ and $W$ over the entire kinematic region. 
Here, the values $-t'=0.132$~GeV$^2$ and $Q^2=1.93$~GeV$^2$ are the mean values of the kinematic variables over the entire kinematic region,  
$0.0$~GeV$^2 \le -t' \le 0.40$~GeV$^2$, $1.0$~GeV$^2 \le Q^2 \le 7.0$~GeV$^2$, and  $3.0$~GeV $\le W \le 6.3$~GeV.
The mean value of $W$ over the entire kinematic region is $4.73$ GeV.
 The NPE amplitude ratio without nucleon-helicity flip, $t_{11}^{(1)}$, is the dominant amplitude ratio. Its real and imaginary parts differ from zero by more 
than five standard deviations. As also already known from the previous analysis~\cite{DC-84}, ${\rm{Re}}\{t_{01}^{(1)}\}$ is significantly non-zero. 
In this analysis, the UPE amplitude ratios without nucleon-helicity flip, 
${\rm{Re}}\{u^{(1)}_{11}\}$ and ${\rm{Im}}\{u^{(1)}_{11}\}$, are individually extracted and found to be non\-ze\-ro
with a significance of about four standard deviations of the total uncertainty.  
The  values of $|u^{(1)}_{11}|$ and $|u^{(2)}_{11}|$, with $\sqrt{|u^{(1)}_{11}|^2+|u^{(2)}_{11}|^2}=0.35\pm 0.06$,   
agree with the result $\sqrt{|u^{(1)}_{11}|^2+|u^{(2)}_{11}|^2} \approx 0.40 \pm 0.02$ obtained in the previous HERMES analysis~\cite{DC-84}. 
The extracted values of the amplitude ratios
show that the main contribution to the term $\sqrt{|u^{(1)}_{11}|^2+|u^{(2)}_{11}|^2}$ comes from the amplitude $U^{(1)}_{11}$ 
without nucleon-helicity flip, and in particular they show that $|U^{(1)}_{11}|^2 \gg |U^{(2)}_{11}|^2$.
The amplitude ratios ${\rm{Im}}\{t^{(2)}_{01}\}$, ${\rm{Im}}\{u^{(2)}_{11}\}$, and ${\rm{Im}}\{u^{(2)}_{10}\}$
deviate from zero by about two standard deviations, while the other extracted amplitude ratios with nucleon-helicity flip 
are consistent with zero within two standard deviations.
The amplitude ratios ${\rm{Im}}\{t^{(2)}_{01}\}$ and ${\rm{Im}}\{u^{(2)}_{10}\}$ are part of those ratios in Eq.~(\ref{prop-t0}), 
which can be nonzero at  $-t'=0$. Among the amplitude ratios that can be zero at  $-t'=0$, only the amplitude ratio ${\rm{Im}}\{u^{(2)}_{11}\}$, which is proportional 
to $\sqrt{-t'}$ at $-t' \to 0$, differs from zero by about two standard deviations of the total uncertainty.

\subsection{Comparison of calculated SDMEs with directly extracted SDMEs}

\begin{figure*}\centering
\includegraphics[width=12.0cm]{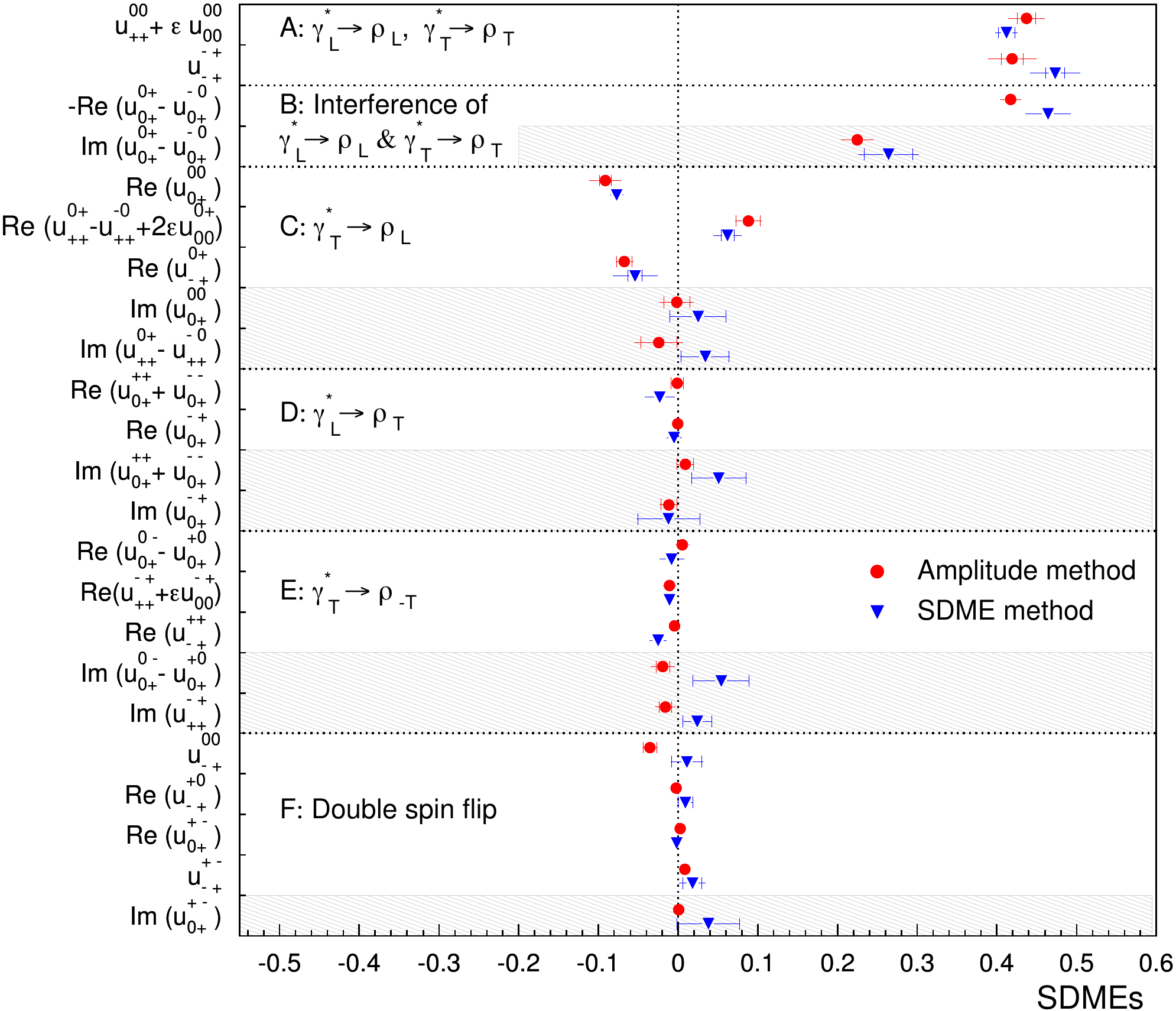}
 \caption{ 
Comparison of the SDMEs $u^{\lambda_V \lambda'_V}_{\lambda_{\gamma}\lambda'_{\gamma}}$ obtained from the amplitude method (red circles)
and from the SDME method (blue triangles). The SDMEs are extracted in the entire kinematic region. 
For the amplitude method a 25-parameter fit is used, while the results of the SDME method are taken from the HERMES data in 
Ref.~\cite{DC-24}.
The points in the shaded area show SDMEs that can be obtained only if the beam is longitudinally polarized. 
An additional scale uncertainty of 2\% originating from the uncertainty on the beam polarization is present for these SDMEs, but not shown.
The inner (outer) error bars represent the statistical (total) uncertainty. 
The SDMEs are ordered according to the SDME classes proposed in Refs.~\cite{DC-24,Ami-th}.}
\label{rel-u-ent}
\end{figure*}

\begin{figure*}\centering
\includegraphics[width=12.0cm]{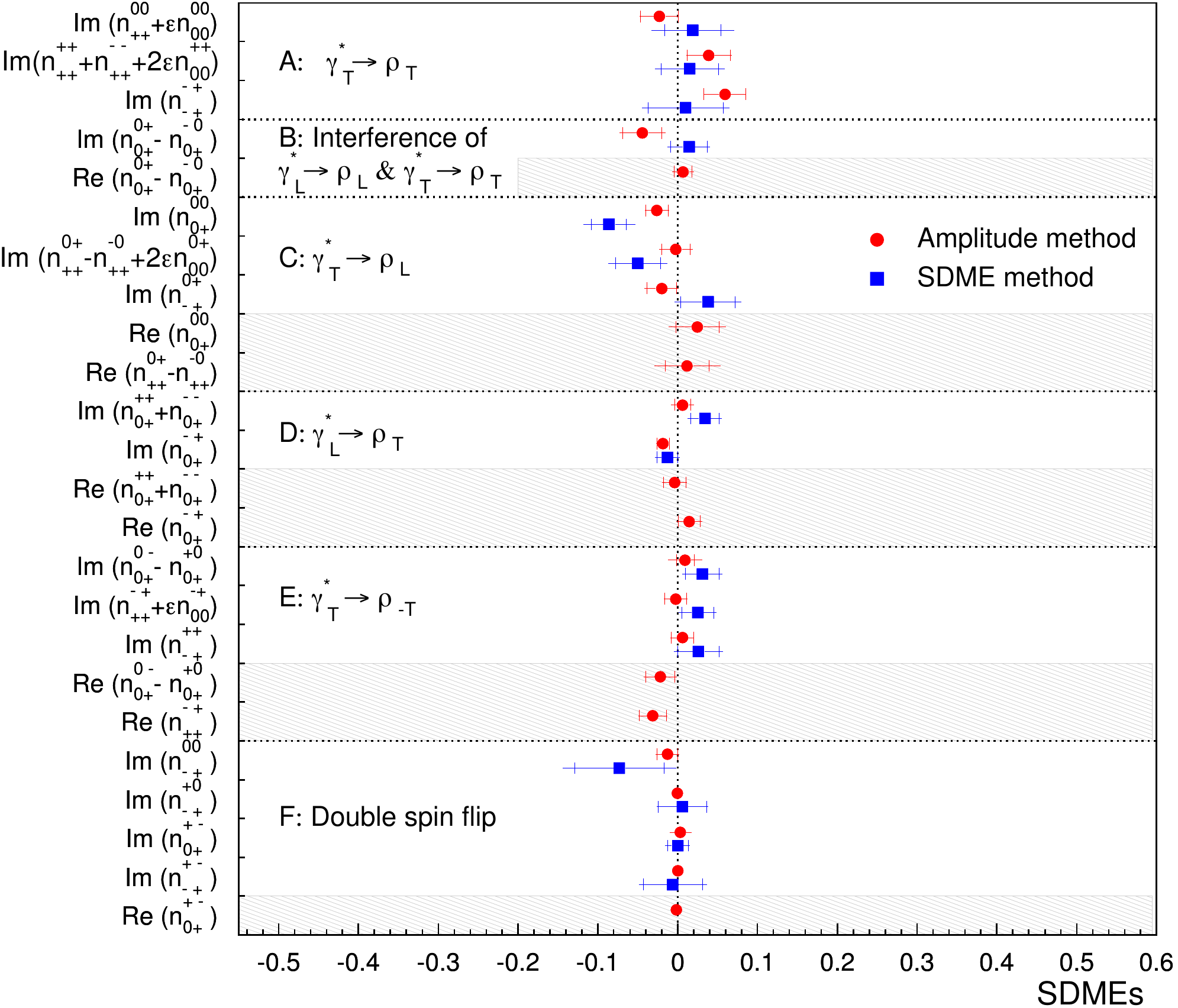}
\caption{ 
Comparison of the SDMEs $n^{\lambda_V \lambda'_V}_{\lambda_{\gamma}\lambda'_{\gamma}}$ obtained from the amplitude method (red circles)
and from the SDME method (blue squares). The SDMEs are extracted in the entire kinematic region. 
For the amplitude method a 25-parameter fit is used, while the results of the SDME method are taken from the HERMES data in 
Ref.~\cite{DC-71}.
The points in the shaded area show SDMEs that can be obtained only  if the beam is longitudinally polarized in addition to the transverse target 
polarization required for all SDMEs here. 
 An additional scale uncertainty of 2\% originating from the uncertainty on the beam polarization is present for these SDMEs, but not shown. The inner (outer) 
error bars represent the statistical (total) uncertainty. An additional scale uncertainty of 8\% originating from the 
uncertainty on the target polarization is present, but not shown. 
The SDMEs are ordered according to the SDME classes proposed in Refs.~\cite{DC-24,Ami-th}.}
\label{rel-n-ent}
\end{figure*}

\begin{figure*}\centering 
\includegraphics[width=12.0cm]{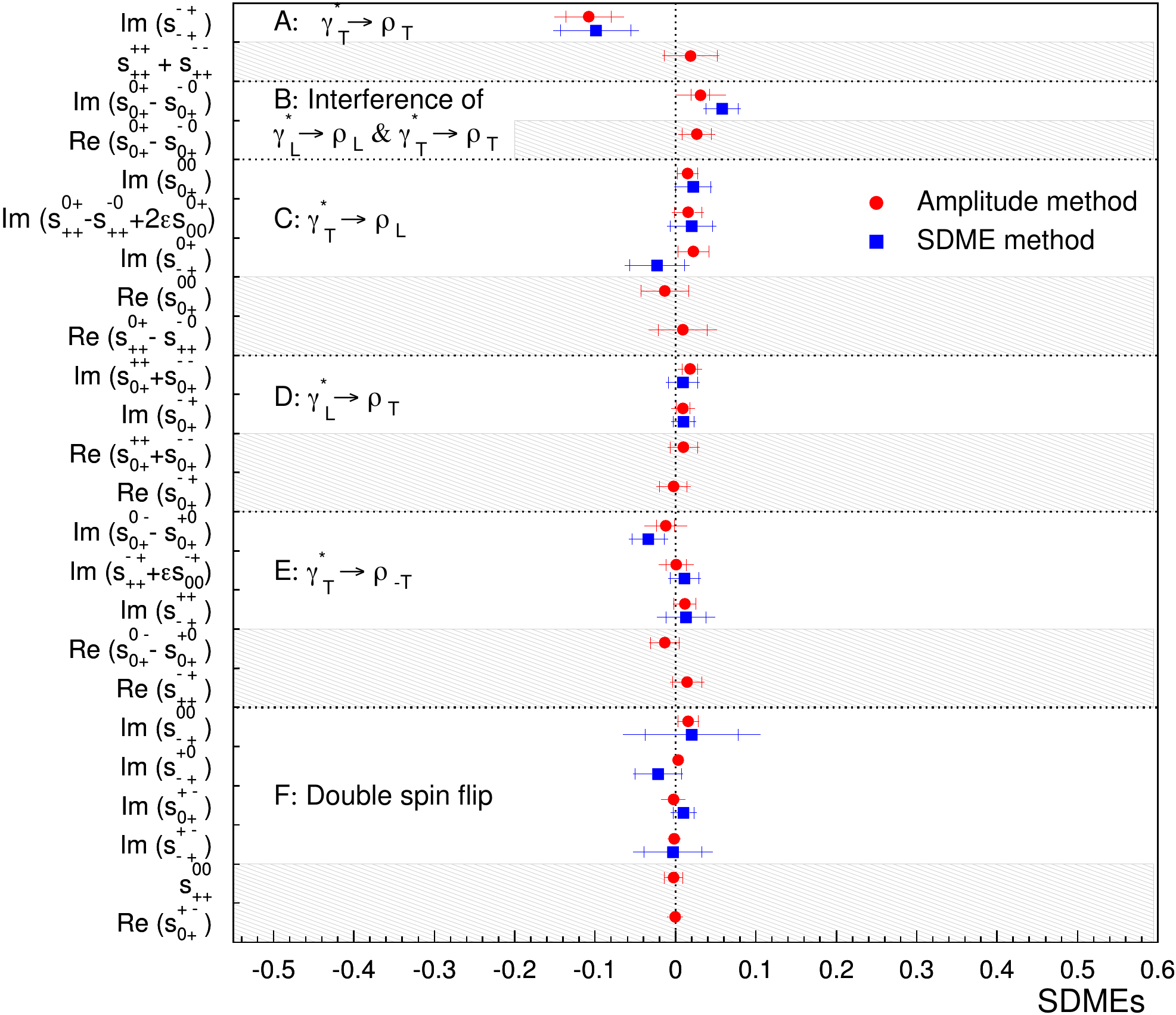}
 \caption{
Comparison of the SDMEs $s^{\lambda_V \lambda'_V}_{\lambda_{\gamma}\lambda'_{\gamma}}$ obtained from the amplitude 
method (red circles)
and from the SDME method (blue squares). The SDMEs are extracted in the entire kinematic region. 
For the amplitude method a 25-parameter fit is used, while the results of the SDME method are taken from the HERMES data in 
Ref.~\cite{DC-71}.
The points in the shaded area show SDMEs that can be obtained only  if the beam is longitudinally polarized in addition to the transverse target 
polarization required for all SDMEs here.  
An additional scale uncertainty of 2\% originating from the uncertainty on the beam polarization is present for these SDMEs, but not shown. 
The inner (outer) error bars represent the statistical (total) uncertainty. An additional scale uncertainty of 8\% originating from the 
uncertainty on the target polarization is present, but not shown. 
The SDMEs are ordered according to the SDME classes proposed in Refs.~\cite{DC-24,Ami-th}.}
\label{rel-s-ent}
\end{figure*}

A comparison of the SDMEs obtained from the SDME method 
in Refs.~\cite{DC-24} and~\cite{DC-71} to those calculated from the amplitude ratios extracted in 
the present analysis is  presented in Figs.~\ref{rel-u-ent}-\ref{rel-s-ent}. 
The SDMEs are calculated in each individual bin using the average kinematics in the parameterizations obtained for the amplitude ratios. 
Furthermore, their mean value is then determined by weighting the SDME value calculated in a given bin by the number of events in this bin.
The correlation matrix for the 25 parameters 
is taken into account for the calculation of the statistical uncertainties of the   
 SDMEs  $u^{\lambda_V \lambda'_V}_{\lambda_{\gamma}\lambda'_{\gamma}}$, $n^{\lambda_V \lambda'_V}_{\lambda_{\gamma}\lambda'_{\gamma}}$, and $s^{\lambda_V 
\lambda'_V}_{\lambda_{\gamma}\lambda'_{\gamma}}$ obtained in the amplitude method.
 As already mentioned in Section~\ref{sec:intro}, the SDMEs $n^{\lambda_V \lambda'_V}_{\lambda_{\gamma}\lambda'_{\gamma}}$ and $s^{\lambda_V 
\lambda'_V}_{\lambda_{\gamma}\lambda'_{\gamma\
}}$, presented in Figs.~\ref{rel-n-ent} and~\ref{rel-s-ent}, can only be extracted from measurements with a transversely polarized target 
so that the helicity-flip amplitude ratios $t^{(2)}_{\lambda_V \lambda_{\gamma}}$ and $u^{(2)}_{\lambda_V \lambda_{\gamma}}$ 
are extracted in this paper for the first time.
The systematic uncertainties of the SDMEs from the amplitude method are determined in an analogous way as for the amplitude ratios by 
varying the relevant parameters, as explained in Sections~\ref{sec-sys_ex} to~\ref{sec-imt11}, and recalculating the 
corresponding SDMEs. 
The total uncertainty is the sum in quadrature of the statistical and the total systematic uncertainties. 

The SDMEs in Figs.~\ref{rel-u-ent}-\ref{rel-s-ent} are reordered according to the SDME classes proposed in Refs.~\cite{DC-24,Ami-th}. 
In these figures also class-F SDMEs are shown. 
Although the double-helicity-flip contribution was a priori not fitted, non-zero values are obtained for this class of SDMEs  
because these SDMEs also receive contributions from other helicity transitions.

Those SDMEs that can be extracted only from data taken with a 
longitudinally polarized lepton beam are shown in shaded areas. 
Figure~\ref{rel-u-ent} shows that for each SDME $u^{\lambda_V \lambda'_V}_{\lambda_{\gamma}\lambda'_{\gamma}}$  determined from our present results, 
there exists an SDME $u^{\lambda_V \lambda'_V}_{\lambda_{\gamma}\lambda'_{\gamma}}$ published in Ref.~\cite{DC-24}.   
 However, Figs.~\ref{rel-n-ent} and~\ref{rel-s-ent} show that for some of the SDMEs $n^{\lambda_V \lambda'_V}_{\lambda_{\gamma}\lambda'_{\gamma}}$ and 
$s^{\lambda_V \lambda'_V}_{\lambda_{\gamma}\lambda'_{\gamma}}$ determined in this analysis
no published results from Ref.~\cite{DC-71} exist, because the beam polarization was 
not exploited in the analyses presented in Ref.~\cite{DC-71}.  
While in Refs.~\cite{DC-24} and~\cite{DC-71} a total of 53 SDMEs could be extracted, the amplitude method presented here 
allows for the calculation of 71 SDMEs based on the extraction of 25 parameters.

As seen from the figures, there is reasonable agreement between SDMEs obtained with the SDME method and those from the amplitude method.
It is possible that the values of the SDMEs obtained in these two methods do not coincide, because the parameter space 
for SDMEs in the SDME method is different from that in the amplitude method. 
Indeed, the SDMEs should belong to a special region in the 71-dimensional real space to give a non-negative angular 
distribution. However, at present the equations determining 
the boundaries of this region are unknown. The physical SDMEs can be represented in terms of 17 helicity-amplitude ratios.    
This restricts the region in the 71-dimensional space. 
This requirement is not taken into account in the SDME method, but it suppresses statistical fluctuations especially when a SDME 
value is  close to the boundary of the allowed region. 
Note that the positivity requirement on the angular distribution is inherent to the amplitude method, while it is not to the 
SDME method, where it is usually imposed artificially. 

\subsection{Comparison to amplitudes calculated in a GPD-based handbag model}

Within the handbag approach (see, e.g., Refs.~\cite{golos3,golos5}), the amplitudes for $\gamma^*_L\to V_L$ and
  $\gamma^*_T\to V_T$ transitions are given by convolutions of appropriate 
  subprocess amplitudes, ${\cal H}_{\lambda_V\lambda'_{q}\lambda_{\gamma}\lambda_q}$, with the GPDs $H^a$, $E^a$, $\tilde{H}^a$, $\tilde{E}^a$,   
 \begin{align}
  F_{\mu \frac{1}{2} \mu \frac{1}{2}} =& \frac{e_0}{2} \sum_{a=u, d, s} e_a \, {\cal C}_{\rho}^a
                    \int \text{d}x \nonumber\\
                 &\;\;  \left[\Big({\cal H}^{a}_{\mu \frac{1}{2} \mu \frac{1}{2}}
                     + {\cal H}^{a}_{\mu -\frac{1}{2} \mu -\frac{1}{2}}\Big)
                      \Big(H^a-\frac{\xi^2}{1-\xi^2}E^a\Big) \right.  \nonumber\\
                 & \;\;\;\; + \!\! \left. \Big({\cal H}^{a}_{\mu \frac{1}{2}\mu \frac{1}{2}} - {\cal H}^{a}_{\mu -\frac{1}{2} \mu -\frac{1}{2}}\Big)
                       \Big(\tilde{H}^a-\frac{\xi^2}{1-\xi^2}\tilde{E}^a\Big)
                       \right]  \nonumber\\
+& \frac{e_0}{2} \, \frac{1}{\sqrt{2}} \int \text{d}x \nonumber\\
                 &\;\; \left[\Big({\cal H}^{g}_{\mu \frac{1}{2} \mu \frac{1}{2}}
                     + {\cal H}^{g}_{\mu -\frac{1}{2} \mu -\frac{1}{2}}\Big)
                      \Big(H^g-\frac{\xi^2}{1-\xi^2}E^g\Big) \right.  \nonumber\\
                 & \;\;\;\; + \!\! \left. \Big({\cal H}^{g}_{\mu \frac{1}{2}\mu \frac{1}{2}} - {\cal H}^{g}_{\mu -\frac{1}{2} \mu -\frac{1}{2}}\Big)
                       \Big(\tilde{H}^g-\frac{\xi^2}{1-\xi^2}\tilde{E}^g\Big)
                       \right]  \nonumber\\
    \label{eq:non-flip}
   \end{align}
   for the case of proton helicity non-flip and
   \begin{align}
   F_{\mu -\frac{1}{2} \mu \frac{1}{2}} =& \frac{e_0}{2} \sum_{a=u,d,s} e_a \, {\cal C}_{\rho}^a \int \text{d}x \nonumber\\
              &\;\;  \left[\Big({\cal H}^{a}_{\mu \frac{1}{2} \mu \frac{1}{2}}+ {\cal H}^{a}_{\mu -\frac{1}{2} \mu -\frac{1}{2}}\Big)E^a \right.\nonumber\\
	   &\;\; \;\;  + \!\! \left. \Big({\cal H}^{a}_{\mu \frac{1}{2} \mu \frac{1}{2}} -  {\cal H}^{a}_{\mu -\frac{1}{2} \mu -\frac{1}{2}}\Big)\,
	            \xi \tilde{E}^a\right]\nonumber\\
   +& \frac{e_0}{2} \, \frac{1}{\sqrt{2}} \int \text{d}x \nonumber\\
  &\;\; \left[\Big({\cal H}^{g}_{\mu \frac{1}{2} \mu \frac{1}{2}}+ {\cal H}^{g}_{\mu -\frac{1}{2} \mu -\frac{1}{2}}\Big)E^g \right.\nonumber\\
	   &\;\; \;\;  +  \!\!  \left. \Big({\cal H}^{g}_{\mu \frac{1}{2} \mu \frac{1}{2}} -  {\cal H}^{g}_{\mu -\frac{1}{2} \mu -\frac{1}{2}}\Big)\,
	            \xi \tilde{E}^g\right]
   \label{eq:flip}		 
   \end{align}
   for the case of proton helicity flip.
 
   Here, $\lambda_{q}$ and $\lambda'_{q}$ are the helicities of the emitted and reabsorbed quarks from the proton, 
   respectively, with $\lambda_{q}=\lambda'_{q}$ 
   for quark helicity non-flip GPDs, $\mu=\lambda_V=\lambda_{\gamma}$, $e_{a}$ are the quark charges in units of the positron charge $e_{0}$, and 
   $x$ is an internal integration variable. 
   The skewness $\xi$ is related to $x_B$ by $\xi=x_B/(2-x_B)$ up to
   corrections of order $1/Q^2$, where $x_B$ is the Bjorken scaling variable defined as 
   \begin{equation}
   x_B = \frac{Q^2}{2\,p\cdot q}  =\frac{Q^2}{2\,M\,\nu}.
   \end{equation} 
   The coefficients ${\cal C}_{\rho}^a$ are appropriate
   flavor factors (${\cal C}_{\rho^0}^u=-{\cal C}_{\rho^0}^d=1/\sqrt{2})$. 
   Because of parity invariance, the first of the each two terms in square brackets in Eqs.~(\ref{eq:non-flip}) and (\ref{eq:flip}) 
   of both the quark and gluon parts
   behave like natural-parity-exchanges (see Eq.~(\ref{npe-amp})), while the 
   second terms are of the unnatural-parity type (see Eq.~(\ref{upe-amp})). 
   For the $\gamma^*_L\to V_L$ transition, there is a rigorous
   proof of factorization in hard subprocesses and GPD 
   amplitudes~\cite{CFS} in the generalized
   Bjorken regime of large $Q^2$, large $W$ but fixed $x_B$. Contributions to longitudinal amplitudes come from 
   GPDs $H$ and $E$, only. In contrast, 
   the $\gamma^*_T\to V_T$ amplitudes are infrared singular in collinear approximation. 
   In order to regularize this singularity, the so-called modified perturbative approach has been used in Refs.~\cite{golos2,golos3} 
   in which quark transverse momenta are retained in the subprocess, while
   the emission and reabsorption of the partons from the proton are still treated
   collinear to the incoming and outgoing proton momenta. The quark transverse
   momenta in the subprocess imply a separation of color charges, which results
   in gluon radiation, as it was calculated in Ref.~\cite{BS}
   in next-to-leading-log approximation and resummed
   to all orders of perturbative QCD. This gluon radiation is also taken into account
   in Refs.~\cite{golos2,golos3}.


  Measurements of the spin asymmetry with a transver\-se\-ly polarized target in exclusive $\pi^+$
  leptoproduction~\cite{Hermes-pi+} revealed that the transversity or helicity-flip GPDs play an important
  role in the $\gamma^*_T\to V_L$ transitions~\cite{golos5}.
  The corresponding amplitudes read
  \begin{eqnarray}
  F_{0-\frac{1}{2} 1 \frac{1}{2}} &=& e_0\sum_{a=u,d,s} e_a \, {\cal C}_{\rho}^a\int dx\, {\cal H}^a_{0-\frac{1}{2} 1 \frac{1}{2}} H_T^a \nonumber,\\
   F_{0\frac{1}{2}\pm 1 \frac{1}{2}}&=& \mp e_0 \frac{\sqrt{-t'}}{4M}\sum_{a=u,d,s} e_a \, {\cal C}_{\rho}^a \int dx \, {\cal H}^a_{0-\frac{1}{2} 1 \frac{1}{2}} 
\bar{E}_T^a\nonumber,\\
  F_{0-\frac{1}{2} -1 \frac{1}{2}} &=& 0.
  \label{eq:tranGPD}
  \end{eqnarray}

  According to the discussion presented in Section~\ref{sec:npe-upe-ampls}, the amplitude $F_{0 \frac{1}{2} \pm 1 \frac{1}{2}}$
  is of natural-parity type, while the proton helicity-flip amplitude $F_{0-\frac{1}{2} 1 \frac{1}{2}}$
  has no specific parity. In the subprocess amplitude ${\cal H}_{0-\frac{1}{2} 1 \frac{1}{2}}$, quark and
  antiquark forming the longitudinally polarized $\rho^0$ meson
  have the same helicity. This fact necessitates the use of a twist-3 meson wave function.

  An important role is played by the pion-pole contribution. 
  As it was discussed in Ref.~\cite{golos4}, the pion pole is to be treated as a 
  one-pion-exchange term, since its evaluation through the GPD
  $\widetilde E$ underestimates the contribution grossly.
  An important element of the pion-pole contribution is the
  $\pi-\rho$ transition form factor $g_{\pi\rho}(Q^2)$. It is estimated 
  to be a third of the $\pi-\omega$ form factor~\cite{golos4} that was extracted
  from the HERMES measurement of the $\omega$ SDMEs~\cite{DC-95}.  
  The factor 1/3 arises from the different 
  quark content of the $\omega$ and $\rho^0$ mesons. 
  The pion pole represents an unnatural-parity contribution and, as can be shown, 
  it contributes dominantly to the amplitudes for transversely polarized vector mesons: 
  \begin{equation}
  F^{pole}_{1\pm \frac{1}{2} \mu \frac{1}{2}} \sim
  \frac{g_{\pi\rho}(Q^2)}{t-m^2_\pi},
  \end{equation}
  with $\mu=0,1$ and $m_\pi$ the mass of the pion.
  The explicit expressions for $F^{pole}$ are given in Ref.~\cite{{golos4}}. They are to be added to the amplitudes from Eqs.~(\ref{eq:non-flip}) and ~(\ref{eq:flip}). 
  The $\gamma_L^{*} \rightarrow V_{T}$ amplitudes receive contributions from only the pion pole.
  With regard to the large $Q^2$-behavior of the $\pi-V$ transition
  form factor, these transition amplitudes are suppressed by $1/Q$ and $1/Q^2$, respectively, 
  as compared to the asymptotically leading $\gamma^*_L\to V_L$ amplitudes. 
  The pion-pole contributions to the $\gamma^*_T\to V_{-T}$ and
  $\gamma^*_T\to V_L$ transition amplitudes are suppressed more strongly 
  and are therefore neglected. As already discussed in Section~\ref{sec:npe-upe-ampls}, 
  there is no UPE contribution for the $\gamma^*_L\to V_L$ amplitudes.

  Details of the calculations of the amplitudes as well as the parametrization of the GPDs, 
  the meson wave-function and the $\pi$-$\rho$ transition form factor can be found in the original papers~\cite{golos3,golos4}.
  The evaluation of the amplitudes
  represent an intermediate step of the calculation of the observables discussed in
  these papers. These amplitudes are divided by $F_{0 \frac12 0 \frac12}=T^{(1)}_{00}$ in order to obtain the amplitude ratios that can be 
  compared to the ones discussed above. The phase convention from Eq.~(\ref{jacobwick}) is taken into account.

\begin{figure*}\centering
\includegraphics[width=10.0cm]{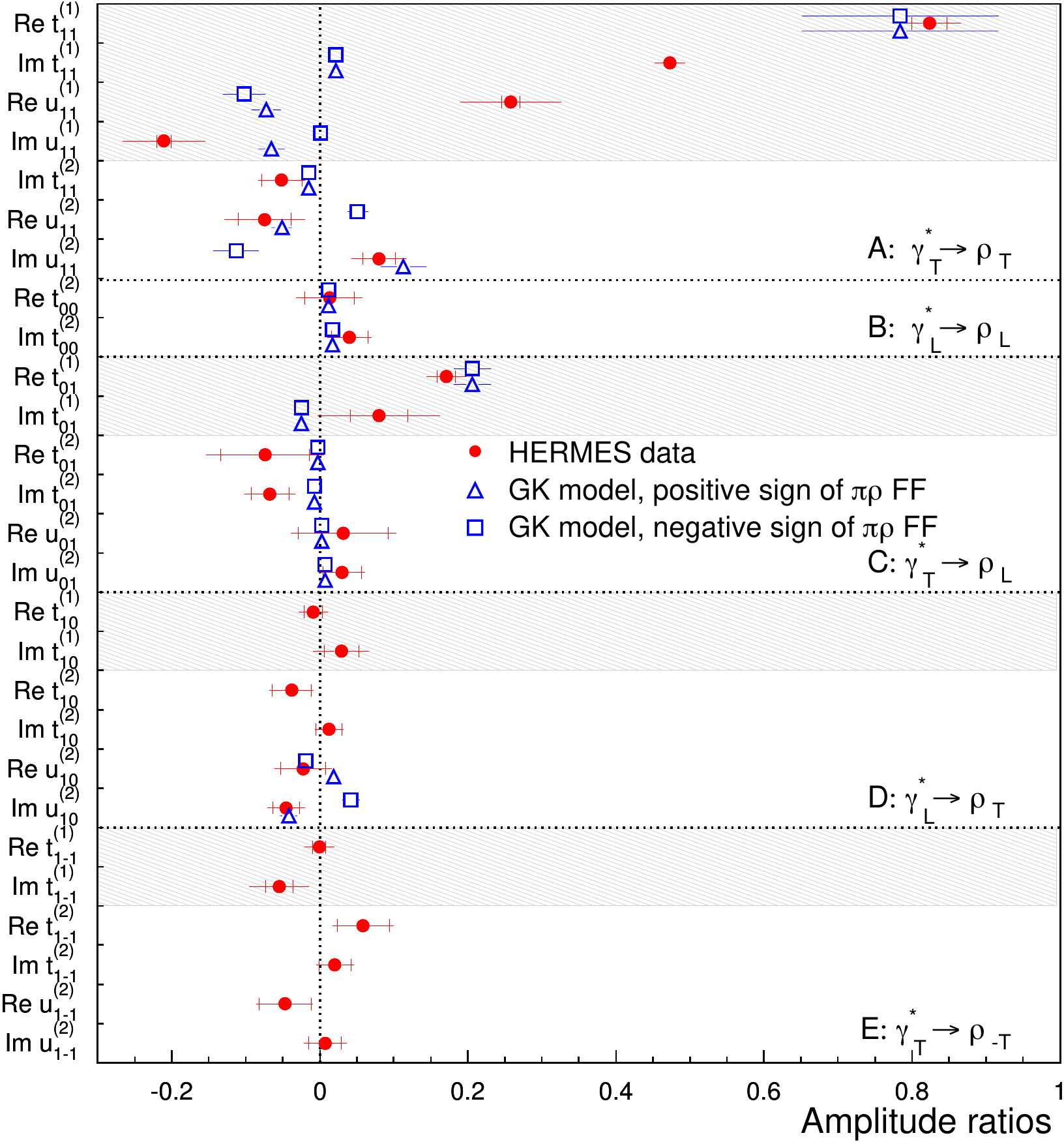}
 \caption{
Comparison of amplitude ratios determined in this paper to those calculated in the GK model, where the phase convention 
from Eq.~(\ref{jacobwick}) is taken into account.
The amplitude ratios that are set to zero in the GK model are not shown.  The inner (outer) error bars represent the statistical (total) uncertainty. 
An additional scale uncertainty of 8\% originating from the uncertainty on the target polarization is present 
for the ratios $t^{(2)}_{\lambda_V \lambda_{\gamma}}$ and $u^{(2)}_{\lambda_V\lambda_{\gamma}}$, but not shown. 
An extra scale uncertainty of 2\% originating from the uncertainty on the beam polarization is present
for the ratios 
${\rm{Im}}\{t^{(1)}_{\lambda_V\lambda_{\gamma}}\}$, ${\rm{Re}}\{t^{(2)}_{\lambda_V\lambda_{\gamma}}\}$, and
${\rm{Re}}\{u^{(2)}_{\lambda_V\lambda_{\gamma}}\}$, but also not shown. 
The amplitude ratios are ordered according to the classes proposed in Refs.~\cite{DC-24,Ami-th}.  The red, filled circles correspond to the 
extracted amplitude ratios and the blue, open triangles (squares) 
represent the result of the GK model calculation using the positive (negative) sign of the $\pi-\rho$ transition form factor.}
\label{amplfit-gk}
\end{figure*}

Figure~\ref{amplfit-gk} shows the comparison of all amplitude ratios determined by fitting the HERMES data to those calculated
using the GK model. The HERMES data are obtained as explained in Section~\ref{sec:resul_disc}, while the values obtained in the GK model are calculated at 
$W=4.73$~GeV, $Q^2=1.93$~GeV$^2$, and $-t'=0.132$~GeV$^2$.
As shown by the following detailed comparison,
good overall agreement is found.

\begin{itemize}
\item[$t_{11}^{(1)}$:] Contributions come from GPDs $H-\xi^2/(1-\xi^2) E \simeq H$. Good agreement is observed for the real part, which is by far the largest 
amplitude ratio. The calculated imaginary part appears to be too small. 
Note that a part of this difference is due to the known underestimation of the relative phase between the $\gamma^*_T\to V_{T}$ and $\gamma^*_L\to V_{L}$ 
amplitudes in the GK model~\cite{DC-24}.
 \item[$u_{11}^{(1)}$:] Contributions come from GPDs $\widetilde{H}$ and the pion pole. The GK calculations underestimate the unnatural-parity contribution to 
 the $\gamma_T^*\to V_T$ amplitude, which is related to the small unnatural-parity cross section used in the GK model~\cite{golos4}. It may be traced back to the 
neglect of the non-pole contribution of the GPD $\widetilde E$ 
or to a too small value for the $\pi-\rho$ transition form factor in the GK model. 
\item[$t_{11}^{(2)}$:] Contributions come from GPDs $E$. The calculated imaginary part agrees with the measurement. 
 \item[$u_{11}^{(2)}$:] In GK calculations only the pion pole contributes since $\widetilde{E}$ is neglected, so that the GK result is mirror symmetric upon sign 
change of the $\pi-\rho$ transition form factor. Good agreement with the data is seen for the  positive sign.
\item[$t_{00}^{(2)}$:] Contributions come from GPDs E. Agreement is observed with the measurement.
\item[$t_{01}^{(1)}$:] Contributions come from GPDs $\bar{E}_T$. Agreement is observed with the measurement.
\item[$t_{01}^{(2)}$:] Contributions from GPDs $H_T$. There is no pion-pole contribution to this ratio, hence data cannot decide on the sign of the form factor. 
The measured imaginary part seems to be lower than the GK calculation.
\item[$u_{01}^{(2)}$:] Contributions from GPDs $H_T$. Since these GPDs have no specific parity, $u_{01}^{(2)}$ is equal to $-t_{01}^{(2)}$ in the GK calculation. 
 \item[$u_{10}^{(2)}$:] Contributions come from the pion pole only, so that the GK result is mirror symmetric upon sign change of the $\pi-\rho$ transition form 
factor. The positive sign is favored by the data.
\end{itemize}

 The $\gamma^*_T\to V_{-T}$ amplitudes, corresponding to the amplitude ratios $t_{1-1}^{(1)}$, $t_{1-1}^{(2)}$, and $u_{1-1}^{(2)}$, are neglected in the GK 
model. This is seen to be in reasonable agreement with the data. 
 Only gluon trans\-ver\-si\-ty GPDs could contribute and the contribution from the pion pole is suppressed by $1/Q^3$ as compared to the longitudinal amplitudes. 
Both are neglected in the GK model. 

As discussed in Section~\ref{sec:choice}, the ratios $u_{01}^{(1)}$, $u_{10}^{(1)}$ 
  and $u_{1-1}^{(1)}$ cannot be determined experimentally in the present analysis and are hence put equal to zero. 
  In the GK model, $u_{01}^{(1)}$ and $u_{1-1}^{(1)}$ are also set equal to zero, while  
  $u_{10}^{(1)}$ is non-zero due to a contribution from the pion pole, but small.
Apart from the $\gamma^*_T\to V_{-T}$ amplitudes, $u_{01}^{(1)}$ and $u_{1-1}^{(1)}$, also $t_{10}^{(1)}$ and $t_{10}^{(2)}$ are set equal to zero.
This is consistent with what is extracted from the data.

As the unnatural-parity amplitudes depend on the sign of the $\pi-\rho$
transition form factor, a conclusion on the sign of the latter can be drawn when comparing the calculated GK amplitude ratios to the data. 
Only the amplitude ratios $u_{11}^{(2)}$ and $u_{10}^{(2)}$ appear sensitive to the sign of the form factor
 and are hence used to calculate the $\chi^2$ per degree of freedom, i.e., $ndf=4$. For the positive sign $\chi^2/ndf=1.8/4$ is obtained and for the negative 
sign $\chi^2/ndf=30.3/4$.
Hence the positive sign of this form factor is clearly favored.

\section{Summary and conclusions}
\label{sec:summary}

Exclusive electroproduction of $\rho^0$ mesons is studied in the HERMES experiment, using data collected with a $27.6$~GeV
 longitudinally polarized electron/positron beam and a transversely polarized hydrogen target in the 
 kinematic region $1.0$~GeV$^2 <Q^2<7.0$~GeV$^2$, $3.0$~GeV $<W<6.3$~GeV, and $-t^\prime<0.4$~GeV$^2$. 
The fit to these data using an unbinned maximum-likelihood method with 25 free parameters permits the extraction of ratios of  
natural-parity-exchange amplitudes 
$T_{\lambda_V \lambda'_N\lambda_{\gamma}\lambda_N}$
without nucleon-helicity flip ($\lambda'_N=\lambda_N$)  and, for the first time,  both the 
natural-parity-exchange  and unnatural-parity-exchange \\ 
amplitudes ($T_{\lambda_V \lambda'_N\lambda_{\gamma}\lambda_N}$ and $U_{\lambda_V \lambda'_N\lambda_{\gamma}\lambda_N}$) with nucleon-helicity flip 
 ($\lambda'_N \neq \lambda_N$), all obtained relative to the amplitude $T_{0\frac{1}{2}0\frac{1}{2}}$, which is the largest amplitude in the kinematic region of 
$Q^2>1$~GeV here considered.
 In particular, the modulus of the amplitude ratio $U_{1\frac{1}{2}1\frac{1}{2}}/T_{0\frac{1}{2}0\frac{1}{2}}$, the real part of 
 $T_{1\frac{1}{2}1\frac{1}{2}}/T_{0\frac{1}{2}0\frac{1}{2}}$ as well as 
 the real and imaginary parts of  the amplitude ratios $T_{1\frac{1}{2}0\frac{1}{2}}/T_{0\frac{1}{2}0\frac{1}{2}}$, 
$T_{1\frac{1}{2}-1\frac{1}{2}}/T_{0\frac{1}{2}0\frac{1}{2}}$ and $T_{0\frac{1}{2}1\frac{1}{2}}/T_{0\frac{1}{2}0\frac{1}{2}}$ 
are extracted. They were also obtained  in the previous HERMES analysis \cite{DC-84} and the amplitude ratios  are in a good agreement with one another.
The values of ${\rm{Im}}\{T_{1\frac{1}{2}1\frac{1}{2}}/T_{0\frac{1}{2}0\frac{1}{2}}\}$ and the phase of 
the ratio $U_{1\frac{1}{2}1\frac{1}{2}}/T_{0\frac{1}{2}0\frac{1}{2}}$ are taken 
from the HERMES results \cite{DC-24,DC-84,HERL1,HERL2}. By performing the fit, the ratios of small nucleon-helicity-flip 
natural-parity-exchange amplitudes 
\begin{align}
T_{1\frac{1}{2}0-\frac{1}{2}},\;
T_{1\frac{1}{2}-1-\frac{1}{2}}, \;
T_{0\frac{1}{2}1-\frac{1}{2}}, \nonumber
\end{align}
and unna\-tu\-ral-parity-exchange amplitudes  
\begin{align}
U_{1\frac{1}{2}1-\frac{1}{2}},\;
U_{1\frac{1}{2}0-\frac{1}{2}},\;
U_{1\frac{1}{2}-1-\frac{1}{2}},\;
U_{0\frac{1}{2}1-\frac{1}{2}}\nonumber
\end{align}
to $T_{0\frac{1}{2}0\frac{1}{2}}$  as well as  
\begin{align}
{\rm{Im}}\{T_{0\frac{1}{2}0-\frac{1}{2}}/T_{0\frac{1}{2}0\frac{1}{2}}\},\;  
{\rm{Im}}\{T_{1\frac{1}{2}1-\frac{1}{2}}/T_{1\frac{1}{2}1\frac{1}{2}}\}, \nonumber\\  
{\rm{Re}}\{T_{0\frac{1}{2}0-\frac{1}{2}}/T_{0\frac{1}{2}0\frac{1}{2}}-T_{1\frac{1}{2}1-\frac{1}{2}}/T_{1\frac{1}{2}1\frac{1}{2}}\} \nonumber
\end{align}
are obtained for the first time, as the data presented here were taken using 
a transversely polarized hydrogen target and a longitudinally polarized lepton beam.

Within the total experimental uncertainty, all determined 
 amplitude ratios with nucleon-helicity flip are consistent with zero.
 The extracted values of the amplitude ratios show that the main contribution to the quantity 
$\frac{\sqrt{|U_{1\frac{1}{2}1\frac{1}{2}}|^2+|U_{1\frac{1}{2}1-\frac{1}{2}}|^2}}{|T_{0\frac{1}{2}0\frac{1}{2}}|}$
obtained in Ref.~\cite{DC-84} originates from the un\-na\-tu\-ral-pa\-ri\-ty-ex\-change amplitude $U_{1\frac{1}{2}1\frac{1}{2}}$ and that 
$|U_{1\frac{1}{2}1\frac{1}{2}}|^2 \gg |U_{1\frac{1}{2}1-\frac{1}{2}}|^2$.
 Furthermore, it is shown that the 53 SDMEs extracted in Refs.~\cite{DC-24,DC-71} can be described with good accuracy  using the 25 amplitude ratios obtained in 
 the present analysis. By also exploiting the longitudinal beam polarization, 18 additional $\rho^0$ SDMEs are determined from the extracted 25 parameters for 
the first time.

The unnatural-parity amplitudes depend on the sign of the $\pi-\rho$
transition form factor, so that the comparison of certain amplitude ratios to calculations within a GPD-based 
handbag model taking into account the contribution from pion exchange
allows the conclusion that the positive sign of this form factor is favored.

Together with precise data on the unpolarized differential cross section $\text{d}\sigma/\text{d}t$ of 
exclusive $\rho^0$ production in deep-inelastic scattering, 
the extracted amplitude ratios could be used to obtain the amplitude $T_{0\frac{1}{2}0\frac{1}{2}}$, for which the factorization property is proven. 

\begin{acknowledgement}
{\bf Acknowledgments~}
We gratefully acknowledge the DESY management for its support and the staff
at DESY aas well as the collaborating institutions for their significant effort.
This work was supported by 
the State Committee of Science of the Republic of Armenia, Grant No. 15T-1C401;
the FWO-Flanders and IWT, Belgium;
the Natural Sciences and Engineering Research Council of Canada;
the Alexander von Humboldt Stiftung,
the German Bundesministerium f\"ur Bildung und Forschung (BMBF), and
the Deutsche For\-schungs\-gemeinschaft (DFG);
the Italian Istituto Nazio\-na\-le di Fisica Nucleare (INFN);
the MEXT, JSPS, and G-COE of Japan;
the Dutch Foundation for Fundamenteel Onderzoek der Materie (FOM);
the Russian Academy of Science and the Russian Federal Agency for 
Science and Innovations;
the Basque Foundation for Science (IKERBASQUE) and MINECO (Juan de la Cierva), Spain;
the U.K.~Engineering and Physical Sciences Research Council, 
the Science and Technology Facilities Council,
and the Scottish Universities Physics Alliance;
as well as the U.S. Department of Energy (DOE) and the National Science Foundation (NSF).
\end{acknowledgement}

\begin{appendix}

\section{The longitudinal double-spin asymmetry}
\label{ap_all}

 The phase $\delta_{u}$ can be determined using the HERMES data~\cite{HERL1,HERL2} taken with a longitudinally polarized hydrogen target with better accuracy 
than using measurements with a 
transversely polarized hydrogen target. The longitudinal double-spin asymmetry in exclusive $\rho^0$-meson electroproduction is defined as~\cite{HERL1} 
\begin{eqnarray}
\nonumber
 A_1^{\rho} & = & 
\frac{\text{d}\sigma_{\frac{1}{2}}/\text{d}t-\text{d}\sigma_{\frac{3}{2}}/\text{d}t}{\text{d}\sigma_{\frac{1}{2}}/\text{d}t+\text{d}\sigma_{\frac{3}{2}}/\text{d}t}\\
 & \approx & 
\frac{|F_{1\frac{1}{2}1\frac{1}{2}}|^2-|F_{1-\frac{1}{2}1-\frac{1}{2}}|^2}
{|F_{1\frac{1}{2}1\frac{1}{2}}|^2+|F_{1-\frac{1}{2}1-\frac{1}{2}}|^2}
\equiv\frac{2{\rm{Re}}{\{t^{(1)}_{11}u^{*(1)}_{11}}\}}{|t^{(1)}_{11}|^2+|u^{(1)}_{11}|^2}.
\label{long-asym}
\end{eqnarray}
 Here, $\text{d}\sigma_{\frac{1}{2}}/\text{d}t$ and $\text{d}\sigma_{\frac{3}{2}}/\text{d}t$ denote the differential cross section for $\rho^0$-meson production 
with a transverse virtual photon, where $1/2$
and $3/2$ are the total projections of the spins of $\gamma^*$ and $p$ onto the photon momentum in the $\gamma^*p$ CM system, respectively.
For the transformations in Eq.~(\ref{long-asym}), Eqs.~(\ref{decomp}), (\ref{symmnat}-\ref{symmunn}) and  
(\ref{def-tjk}-\ref{def-ujk}) are used.
Equation~(\ref{long-asym}) can be rewritten in terms of the phase $\delta_{u}$ and the phase $\delta$ of the amplitude ratio 
$t^{(1)}_{11}$ (which is nothing else than the 
phase difference between the amplitudes $T^{(1)}_{11}$ and $T^{(1)}_{00}$) as
\begin{eqnarray} 
A_1^{\rho}=\frac{2|t^{(1)}_{11}||u^{(1)}_{11}| \cos(\delta_{u}-\delta)}{|t^{(1)}_{11}|^2+|u^{(1)}_{11}|^2}.
\label{a1rho}
\end{eqnarray} 
 Since the moduli of the amplitude ratios $|t^{(1)}_{11}|$ and $|u^{(1)}_{11}|$ and the phase $\delta$ were extracted from HERMES data~\cite{DC-24,DC-84}, the 
measured value of the asymmetry $A_1^{\rho}=0.24$ \cite{HERL1,HERL2} gives a 
value of $\delta_{u}-\delta \approx \pm 69.8$ degrees. The sign of the latter is obtained from 
\begin{eqnarray}
\frac{U^{(1)}_{11}}{T^{(1)}_{11}}=\Bigl | \frac{U^{(1)}_{11}}{T^{(1)}_{11}}\Bigr | \exp \{ i(\delta_{u}-\delta)\}.
\label{ratu11t11}
\end{eqnarray}
The UPE amplitude $U^{(1)}_{11}$ is mainly the amplitude of pion exchange, 
which is a real positive function at small $-t'$. At high energies, the imaginary part of the 
NPE amplitude $T^{(1)}_{11}$ is positive and much larger than its real part, so that the ratio
$U^{(1)}_{11}/T^{(1)}_{11}$  has to have a negative imaginary part leading to 
 $\sin (\delta_{u}-\delta) <0$. This gives $\delta_{u}-\delta \approx -69.8$ degrees and hence $\delta_{u}=-39.2$ degrees if the value $\delta=30.6$ degrees is 
taken from Ref.~\cite{DC-24}. The range of $\delta_{u}$
considering the values 0.38 and 0.1 for the asymmetry $A_1^{\rho}$ (corresponding to $\pm $ one standard deviation) is
 $\delta_{u}=-26.2$ and $\delta_{u}=-51.1$ degrees.

\section{Linear contribution of amplitudes $T^{(2)}_{00}$ and $T^{(2)}_{11}$ to the angular distribution}
\label{sec-8.3}

Let us use the following parameterization for the ratios of the small amplitudes $T^{(2)}_{00}$ and $T^{(2)}_{11}$ to the big amplitudes
$T^{(1)}_{00}$ and $T^{(1)}_{11}$:
\begin{eqnarray}
\frac {T^{(2)}_{00}}{T^{(1)}_{00}}=\beta_{00}+i\alpha_{00},
\label{rat-t00}\\
\frac {T^{(2)}_{11}}{T^{(1)}_{11}}=\beta_{11}+i\alpha_{11}
\label{rat-t11}.
\end{eqnarray}
Hence the terms in the angular distribution that contain linear contributions from 
the small amplitudes $T^{(2)}_{00}$ and $T^{(2)}_{11}$ can be written as~\cite{Diehl}
\begin{align}
 \Delta W & (\Phi,\phi_s, \theta,\phi)   \nonumber \\
 =& \; \frac {3P_T\sin(\Phi+\phi_s)}{8 \pi^2\mathcal{N}}   \Bigl \{2\alpha_{00}\epsilon|T^{(1)}_{00}|^2\cos^2 \theta \nonumber \\
& \qquad +\alpha_{11}|T^{(1)}_{11}|^2\sin^2 \theta \, [1+\epsilon \cos(2\Phi-2\phi)] \nonumber \\
& \qquad -\sqrt{2}\sin \theta \cos \theta \Bigl [ \cos (\Phi-\phi)\sqrt{\epsilon(1+\epsilon)} \nonumber \\
& \qquad\qquad \times [(\beta_{11}-\beta_{00}){\rm{Im}}\{T^{(1)}_{00}(T^{(1)}_{11})^*\} \nonumber \\
& \qquad\qquad\quad -(\alpha_{11}+\alpha_{00}){\rm{Re}}\{T^{(1)}_{00}(T^{(1)}_{11})^*\}] \nonumber \\
& \qquad\quad +P_B\sin (\Phi-\phi)\sqrt{\epsilon(1-\epsilon)} \nonumber \\
& \qquad\qquad \times [(\beta_{11}-\beta_{00}){\rm{Re}} \{T^{(1)}_{00}(T^{(1)}_{11})^*\}  \nonumber \\
& \qquad\qquad\quad +(\alpha_{11}+\alpha_{00}){\rm{Im}}\{T^{(1)}_{00}(T^{(1)}_{11})^*\}]\Bigr ]\Bigr \},
\label{deltaW}
\end{align}
where the normalization factor $\mathcal{N}$ is defined by Eqs.~(\ref{ntotal}-\ref{norm-l}).
Here, the azimuthal angle $\phi_s$ between the transverse component of the target polarization 
with respect to the virtual-photon momentum and the lepton scattering plane is defined 
as in Ref.~\cite{Diehl}. 
It is related to the angle $\Psi$ through the following equations:
\begin{eqnarray}
\cos \phi_s=\frac{\cos \theta_{\gamma} \cos \Psi}{\sqrt{1- \sin^2 \theta_{\gamma} \cos^2 \Psi}},
\label{cosphis}\\
\sin \phi_s=\frac{\sin \Psi}{\sqrt{1- \sin^2 \theta_{\gamma} \cos^2 \Psi}},
\label{sinphis}
\end{eqnarray}
where the angle $\theta_{\gamma}$ is defined by Eqs.~(\ref{csthetgam}) and (\ref{snthetgam}). 
Note that the angle $\Phi$ used in Eq.~(\ref{deltaW})   is related to $\phi_{[6]}$ used in Ref.~\cite{Diehl} by the equation
$\Phi=2 \pi - \phi_{[6]}$.
Also, the angle $\phi$ of the present paper is denoted in Ref.~\cite{Diehl} by ${\varphi}$. 

As seen from Eq.~(\ref{deltaW}) there is a contribution from 
the combination $(\beta_{11}-\beta_{00})$ of the parameters $\beta_{11}$ and $\beta_{00}$, but
 there is no contribution from $(\beta_{11}+\beta_{00})$ to the angular distribution. Therefore, the three parameters 
$\alpha_{11}$, $\alpha_{00}$, and
$(\beta_{11}-\beta_{00})/2$ can be extracted from the angular distribution, while 
\begin{eqnarray}
\kappa & = & \frac{1}{2}(\beta_{11}+\beta_{00}) \equiv \frac{1}{2} {\rm{Re}}  \Bigl\{ \frac{T^{(2)}_{11}}{T^{(1)}_{11}}+\frac {T^{(2)}_{00}}{T^{(1)}_{00}} \Bigr\} \\
& \equiv & \frac{1}{2}{\rm{Re}}\Bigl\{\frac{{t}^{(2)}_{11}}{{t}^{(1)}_{11}}+t^{(2)}_{00}\Bigr\}
\label{kappabeta}
\end{eqnarray}
cannot be reliably obtained from data on a transversely polarized target. Note that the second-order contributions proportional  to 
$|\beta_{11}|^2$ and
$|\beta_{00}|^2$ exist, since at least the normalization factor $\mathcal{N}$ in formula (\ref{neumann}) defined by Eqs.~(\ref{ntotal}-\ref{norm-l}) 
contains  squared moduli of 
all amplitudes. However, these second-order contributions are negligibly small.    

\section{Correlation matrix for fit parameters}
\label{sec-8.4}

Table~\ref{corrmatr} lists the correlations between the 25 parameters of the fit of helicity-amplitude ratios defined in Table~\ref{param25}.

\begin{table*}
\caption{\label{corrmatr} The correlation matrix for the 25 parameters defined in Table~\ref{param25}. }
\renewcommand{\arraystretch}{1.3}
\setlength{\tabcolsep}{3pt}
\hspace*{2.0cm}
\begin{sideways}
\begin{tabular}{|l|c|c|c|c|c|c|c|c|c|c|c|c|c|c|c|c|c|c|c|c|c|c|c|c|c|}
\hline
 & $b_1$& $b_2$& $b_3$& $b_4$& $b_5$& $b_6$& $b_7$& $b_8$& $b_9$& $b_{10}$& $b_{11}$& $b_{12}$& $b_{13}$& $b_{14}$& $b_{15}$& $b_{16}$& $b_{17}$& $b_{18}$& 
$b_{19}$& $b_{20}$& $b_{21}$& $b_{22}$& $b_{23}$& $b_{24}$& $b_{25}$ \\
\hline
$b_1$&  1.00 \\
$b_2$& 0.08& 1.00 \\
$b_3$& -0.01&-0.03& 1.00 \\
$b_4$& 0.05& 0.02&-0.20& 1.00 \\
$b_5$&0.07& 0.00&-0.16&-0.19& 1.00 \\
$b_6$&-0.05& 0.02&-0.19&-0.03&-0.20& 1.00 \\
$b_7$&-0.05& 0.02& 0.07& 0.08&-0.39&-0.18& 1.00 \\
$b_8$& -0.01& 0.10& 0.01& 0.01& 0.01& 0.03& 0.01& 1.00 \\
$b_9$& 0.23&-0.05&-0.03&-0.02&-0.02&-0.09&-0.06&-0.21& 1.00 \\
$b_{10}$&0.00& 0.04&-0.03& 0.02& 0.04&-0.19& 0.19& 0.02&-0.01& 1.00 \\
$b_{11}$&0.00& 0.02&-0.06& 0.08&-0.02&-0.02& 0.04& 0.04& 0.00& 0.05& 1.00 \\
$b_{12}$&-0.05&-0.02&-0.01& 0.12& 0.04&-0.04&-0.02& 0.01&-0.13&-0.17& 0.08& 1.00 \\
$b_{13}$&-0.02&-0.01& 0.10&-0.09&-0.07&-0.07& 0.13& 0.01&-0.05& 0.08&-0.33& 0.03& 1.00 \\
$b_{14}$& -0.22& 0.13& 0.05& 0.03&-0.01& 0.06& 0.05&-0.11&-0.57&-0.02&-0.02& 0.09& 0.05& 1.00 \\
$b_{15}$&0.27&-0.11&-0.07&-0.03& 0.00&-0.05&-0.06&-0.14& 0.62& 0.03& 0.01&-0.10&-0.06&-0.71& 1.00 \\
$b_{16}$&0.05&-0.05& 0.02& 0.05&-0.08&-0.13& 0.17&-0.02& 0.12& 0.46& 0.15&-0.26& 0.06&-0.16& 0.21& 1.00 \\
$b_{17}$& 0.06&-0.01& 0.03& 0.04&-0.07&-0.11& 0.13&-0.02& 0.08& 0.49&-0.14&-0.41&-0.03&-0.07& 0.12& 0.42& 1.00 \\
$b_{18}$& 0.00& 0.04& 0.23& 0.04& 0.09&-0.10&-0.02&-0.01& 0.05& 0.31&-0.06&-0.45&-0.11& 0.00&-0.02& 0.17& 0.37& 1.00 \\
$b_{19}$&0.01&-0.03& 0.10& 0.07& 0.02&-0.15& 0.16& 0.01& 0.02& 0.48& 0.06&-0.44& 0.15&-0.01& 0.01& 0.40& 0.35& 0.47& 1.00 \\
$b_{20}$& 0.07&-0.10&-0.03&-0.03& 0.02&-0.06&-0.03&-0.08& 0.20&-0.02&-0.02&-0.02&-0.03&-0.23& 0.28& 0.07& 0.02&-0.05&-0.04& 1.00 \\
$b_{21}$&-0.07& 0.13& 0.06& 0.03&-0.04& 0.10& 0.00& 0.15&-0.05& 0.01& 0.04&-0.01& 0.01& 0.14&-0.27&-0.07&-0.03& 0.10& 0.04&-0.53& 1.00 \\
$b_{22}$&-0.02& 0.03&-0.01&-0.27&-0.23& 0.34&-0.06& 0.01&-0.03&-0.01&-0.09&-0.08& 0.03& 0.05&-0.07&-0.09& 0.04&-0.06&-0.07&-0.07& 0.13& 1.00 \\
$b_{23}$& 0.02&-0.03&-0.25& 0.40&-0.17& 0.09&-0.11&-0.02& 0.08&-0.07& 0.01& 0.04&-0.09&-0.03& 0.03& 0.03&-0.03&-0.01&-0.09& 0.00& 0.07& 0.02& 1.00 \\
$b_{24}$&-0.06&-0.04&-0.09&-0.22&-0.16& 0.31&-0.06&-0.01&-0.03&-0.03& 0.04&-0.06& 0.13& 0.04&-0.02& 0.02&-0.06&-0.19&-0.04&-0.02& 0.03& 0.08&-0.02& 1.00 \\
$b_{25}$& 0.00&-0.02&-0.33& 0.40&-0.21& 0.14&-0.09&-0.02& 0.04& 0.00& 0.01& 0.10&-0.07&-0.03& 0.04& 0.05& 0.01&-0.12&-0.13& 0.00& 0.00& 0.00& 0.37& 0.14& 1.00\\
\hline
 & $b_1$& $b_2$& $b_3$& $b_4$& $b_5$& $b_6$& $b_7$& $b_8$& $b_9$& $b_{10}$& $b_{11}$& $b_{12}$& $b_{13}$& $b_{14}$& $b_{15}$& $b_{16}$& $b_{17}$& $b_{18}$& 
$b_{19}$& $b_{20}$& $b_{21}$& $b_{22}$& $b_{23}$& $b_{24}$& $b_{25}$ \\
\hline
\end{tabular}
\end{sideways}
\end{table*}

\end{appendix}


\end{document}